\documentclass[twocolumn,           
               showpacs,            
               nopreprintnumbers,     
               aps,                 
               prd,          	    
               letterpaper,             
              groupeaddress,      
               nofootinbib,         
               tightenlines,        
               floats,floatfix,      
               showkeys
               ]{revtex4-1}
               
\usepackage[toc,page]{appendix}
\usepackage{graphicx}
\usepackage{dcolumn}
\usepackage{bm}
\usepackage{amsmath}
\usepackage{amsfonts,amssymb}
\usepackage{color}
\usepackage{enumerate}
\usepackage{float}
\usepackage{subfig}
\usepackage{ulem}

\begin{document}

\title{A hybrid model of viscous and Chaplygin gas to tackle the Universe acceleration}

\author{A. Hern\'andez-Almada$^1$}
\email{ahalmada@uaq.mx}

\author{Miguel A. Garc\'ia-Aspeitia$^{2,3}$}

\author{M. A. Rodr\'iguez-Meza$^{4}$}

\author{V. Motta$^5$}

\affiliation{$^1$Facultad de Ingenier\'ia, Universidad Aut\'onoma de Quer\'etaro, Centro Universitario Cerro de las Campanas, 76010, Santiago de Quer\'etaro, M\'exico.}
\affiliation{$^2$Unidad Acad\'emica de F\'isica, Universidad Aut\'onoma de Zacatecas, Calzada Solidaridad esquina con Paseo a la Bufa S/N C.P. 98060, Zacatecas, M\'exico.}
\affiliation{$^3$Consejo Nacional de Ciencia y Tecnolog\'ia, \\ Av. Insurgentes Sur 1582. Colonia Cr\'edito Constructor, Del. Benito Ju\'arez C.P. 03940, Ciudad de M\'exico, M\'exico.}
\affiliation{$^4$Departamento de F\'isica, Instituto Nacional de Investigaciones Nucleares, Apartado Postal 18-1027, 11801 Ciudad de M\'exico, M\'exico.}
\affiliation{$^5$Instituto de F\'isica y Astronom\'ia, Facultad de Ciencias, Universidad de Valpara\'iso, Avda. Gran Breta\~na 1111, Valpara\'iso, Chile.}
\begin{abstract}
Motivated by two seminal models proposed to explain the Universe acceleration, this paper is devoted to study a hybrid model which is constructed through a generalized Chaplygin gas with the addition of a bulk viscosity. We call the model a Viscous Generalized Chaplygin Gas (VGCG) and its free parameters are constrained through several cosmological data like the Observational Hubble Parameter, Type Ia  Supernovae, Baryon Acoustic Oscillations, Strong Lensing Systems, HII Galaxies and using Joint Bayesian analysis. In addition, we implement a Om-diagnostic to analyze the VGCC dynamics and its difference with the standard cosmological model. The hybrid model shows important differences when compared with the standard cosmological model. Finally, based on our Joint analysis we find that the VGCG could be an interesting candidate to alleviate the well-known Hubble constant tension. 
\end{abstract}

\date{\today}

\maketitle

\section{Introduction}

The origin of the accelerated expansion of the Universe in recent epochs is believed to be caused by an unknown component of the universe called the Dark Energy (DE). DE is one of the cornerstones in the modern cosmology. Despite that the DE is a strange phenomenon which evolves into an Einstein-D'Sitter structure, it is confirmed by several observations such as the Cosmic Microwave Background Radiation (CMB) \cite{Planck:2018}, the distance modulus of type Ia Supernovae (SnIa) \cite{Riess:1998,Perlmutter:1999,Scolnic:2018} and recently through the large-scale structure studies \cite{LargeScale}. 

The simplest candidate to explain DE comes from adding a cosmological constant (CC with value $\Lambda$) \cite{Carroll:2000}  in the Einstein field equations, i.e., a geometrical correction to them. On the other hand, such CC can also be seen as an effective fluid with an equation of state (EoS) $w=-1$ when it is included through the energy-momentum tensor. 
Nevertheless, the CC origin as quantum vacuum fluctuations is not well understood \cite{Zeldovich,Weinberg}, encouraging the necessity of new approaches to understand its nature. Motivated by the CC problems, the community has been exploring alternatives to the CC which can be classified in two kinds of solutions: modify the Einstein field equations or include effective fluids to them. In this sense, a zoo of DE models has been appearing in the literature. We can list for instance, models related to modified gravity as
$f(R)$ theories \cite{DeFelice:2010aj,Jaime:2010kn,Jaime:2012nj}, 
the extra dimensions \cite{Maartens:2010ar,Garcia-Aspeitia:2018fvw}, 
unimodular gravity \cite{Josset:2016,Garcia-Aspeitia:2019yni,Garcia-Aspeitia:2019yod,Corral:2020lxt}. Conversely, models associated to effective fluids as DE are  
phenomenological dark energy which is related with emergent DE models \cite{Li_2019,Li:2020ybr,Hernandez-Almada:2020uyr}, 
quintessence and phantom DE \cite{Copeland:2006wr},
dissipative fluids \cite{Cruz:2019wbl,Cruz:2017bcv,Brevik:2005bj},
Chaplygin gas (CG) \cite{Chaplygin,Kamenshchik:2001cp}, Generalized Chaplygin Gas (GCG) \cite{Kamenshchik,Bilic,Fabris}, 
and like-Chaplygin gas \cite{Hernandez-Almada:2018osh}. 
For a review of these models see for instance \cite{Copeland:2006wr}.

An interesting property of Chaplygin fluids is that they are able to describe Dark Matter (DM) and DE as an unique perfect fluid through an EoS of the form $p_c=-A\rho_c^{-\alpha}$, where $0\leq\alpha\leq1$ and $A>0$, where $p_c$ and $\rho_c$ are pressure and density respectively. This kind of models have been analyzed for inflationary epochs through the formalism of Hamilton-Jacobi \cite{Villanueva_2015}, and later  were studied at late epochs \cite{Kamenshchik,Bilic,Fabris,Hernandez-Almada:2018osh}. In spite of the advantages, this kind of models experience several problems such as oscillations and exponential blowup in the DM power spectrum, in disagreement with the observations \cite{Sandvik:2004}, which in turn affects the formation of structures. Additionally, several authors \cite{Perrotta:2004} have shown that the trajectories in the context of dynamical systems are unstable, inducing a fine tuning in the initial condition.

On the other hand, non perfect fluids are another approach to tackle the Universe acceleration. In fact, they also allow the description of DM and DE in the context of unified fluids through the inclusion of kinds of viscosity, bulk and shear. However, the bulk viscosity is the best candidate to face the DE problem because satisfies the cosmological principle and it could be studied through two main approaches: Eckart formalism \cite{Eckart} and Israel-Steward (IS) theory \cite{Israel}. The main difference between them is that the Eckart formalism is a non causal theory and, as a consequence, it is simpler than the IS. Indeed, only one bulk viscosity model of the form $\xi \sim \rho^{s}$ (where $\rho$ is the energy density considered typically as dust matter) has been studied \cite{Cruz,CRUZ2017159,Cruz:2018psw,CruzyHernandez,Lepe} in the IS theory, while several models have been explored \cite{Murphy,Padma,Brevik,Xin_He,Avelino, AlmadaViscoso, Folomeev, Almada:2020} in the Eckart one.

In this paper, we present a hybrid model which consists of a fusion of viscous effects and the features of Chaplygin gas, hereafter Viscous Generalized Chaplygin Gas (VGCG), to tackle the late accelerated expansion of the Universe. Naturally, VGCG is also under the unified dark fluids approach as GCG and viscous fluids. This kind of hybrid models was originally proposed by \cite{ZHAI:2006} and are able to avoid causality problems that arise when only dissipative fluids are considered. This model has been studied by several authors \cite{Xu:2012, Saadat2:2013, Li-Xu:2013, Saadat:2013, Jawad:2016dcp, Szydlowski:2020ilx}. For instance, \cite{Jawad:2016dcp} studied the spherical top-hat collapse both in Einstein gravity and in loop quantum cosmology. Additionally, cosmological bounds of the free parameters were established by \cite{Li-Xu:2013}, finding that the viscosity terms ($\xi_0=7.08\times10^{-4}$) may affect the CMB power spectrum mainly on the height of the acoustic peaks related to the matter density but consistent to those obtained by $\Lambda$CDM.
Additionally, one of the most important points to propose this kind of hybrid models is that they could alleviate the oscillations that cause the blowup in the DM power spectrum for GCG models \cite{Amendola_2003,Sandvik:2004}. The problem lies in the equation for the density fluctuations $\delta_k$, in particular in the nonzero sound speed where, under some regime, the fluctuations become violently unstable and grow exponentially. Based on the results regarding CMB power spectrum for VGCG \cite{Li-Xu:2013}, the viscosity terms in the VGCG could also act as a damping term in the DM power spectrum as occurs in the CMB spectrum ($\xi_0=7.08\times10^{-4}$), reducing such instabilities. Therefore, this kind of models must be studied in detail, starting from the background cosmology.

This paper is devoted to analyze this hybrid model considering the Eckart formalism to add its viscous effects from the cosmographics and thermodynamics point of view. Firstly, we constrain the parameter space of the VGCG by performing a Bayesian analysis and using the following samples: 
observational Hubble parameter (OHD) \cite{Magana:2017nfs}, 
type Ia supernovaes (SnIa) \cite{Scolnic:2018}, 
baryon acoustic oscillations (BAO) \cite{nunes:2020}, 
strong lensing systems (SLS) \cite{Amante:2019xao}, 
and HII Galaxies (HIIG) \cite{Chavez2012,Chavez2014,Terlevich2015,Chavez2016,GonzalezMoran2019}. 
Furthermore, we also perform a Bayesian joint analysis combining the mentioned samples to study the cosmography of the model, $\{q,j\}$ phase-space and a test related to the well-known $Om$-diagnostic. Finally, we discuss the near equilibrium condition imposed by the Thermodynamics to any dissipative fluid.

The manuscript is structured as follow: Sec. \ref{Sec:model} presents the GCG model in presence of viscosity and give general expressions to reconstruct and analyze the cosmography parameters. Sec. \ref{sec:constraints} gives a summary of the cosmological data and presents the methodology to establish bounds on the phase-space of the model parameters. Sec. \ref{sec:Results} discusses the results and finally, Sec. \ref{sec:summary} makes a summary and shows the conclusions.  

\section{Viscous generalized Chaplygin gas revisited} \label{Sec:model}

In a spatially flat Universe described by the Friedman-Lemaitre-Robertson-Walker (FLRW) metric, a generalized Chaplygin gas in presence of viscosity is studied. The model describes the evolution of baryons ($b$) and relativistic species ($r$) as perfect fluids with EoS $w_b=p_b/\rho_b=0$ and $w_r=p_r/\rho_r=1/3$ respectively, where $\rho_{b,r}$ is their energy density. Additionally, it is considered a non perfect fluid following the GCG EoS $p_c = -A\rho_c^{-\alpha}$, where $p_c$ and $\rho_c$ are the pressure and energy density respectively and $A$ and $\alpha$ are appropriate constants. Based on Eckart formalism, the viscous effects are introduced in the energy-momentum tensor $T^{\mu\nu}= \rho_t U^\mu U^\nu + p_{eff}( g^{\mu\nu} + U^\mu U^\nu)$ as an effective pressure, $p_{eff}$, where $U^{\mu}$ is the four-velocity. In this context, the total effective pressure is the sum of the barotropic pressure of the fluids and the viscous pressure, 
\begin{equation} \label{eq:eoseff}
    p_{eff} = \sum_i p_i + \Pi\,, 
\end{equation}
where $\Pi= -3\xi H$, $\xi$ is the bulk viscosity, $p_i$ ($i=b,r,c$) is the pressure of the fluid and $H \equiv \dot{a}/a$ is the Hubble parameter. Then the Friedmann and continuity equations are
\begin{eqnarray}
    H^2 &=& \frac{\kappa^2}{3}\left( \rho_b + \rho_r + \rho_c \right), \\
    \dot{\rho_b} &=& - 3 \rho_b H, \\
    \dot{\rho_r} &=& - 4 \rho_r H, \\
    \dot{\rho_c} &=& - 3 (\rho_c + \Tilde{p}_c) H \,,
\end{eqnarray}
where $\kappa^2\equiv8\pi G$ and dots are derivatives with respect to cosmic time. By integrating the first two continuity equations, we have $\rho_i = \rho_{i0}a^{-3(w_i+1)}$ where $\rho_{i0}$ is the initial energy density, for $i=b,r$. To integrate the one for the VGCG we will assume that $\xi = \xi_0 \rho_c^{1/2}$, being $\xi_0$ an appropriate constant and $H^2\approx \kappa^2 \rho_c/3$ (a good approximation for the late stage of the Universe), thus we obtain
\begin{equation}
    \rho_{c} = \rho_{c0} \left[ B_s  + (1-B_s)a^{-3(1+\alpha)(1-\sqrt{3}\xi_0)} \right]^{\frac{1}{1+\alpha}}\,,
\end{equation}
where $B_s=A_s(1-\sqrt{3}\xi_0)^{-1}$ and $A_s = A/\rho_{c0}^{1+\alpha}$. Hence, the dimensionless Hubble parameter $E(z) \equiv H(z)/H_0$ in terms of the redshift $z$ is given by
\begin{eqnarray}
    E(z)^2 &=& \Omega_{c0}\left[ B_s  + (1-B_s)(1+z)^{-3(1+\alpha)(1-\sqrt{3}\xi_0)} \right]^{\frac{1}{1+\alpha}} \nonumber \\
    &+& \Omega_{b0}(1+z)^3 + \Omega_{r0}(1+z)^4 \label{eq:E2}
\end{eqnarray}
where we have used the relation $a=(1+z)^{-1}$ and $\Omega_{c0}=1-\Omega_{b0}-\Omega_{r0}$ which comes from flatness condition $E(0)=1$. It is straightforward that, for $\xi_0\rightarrow 0$, $B_s\rightarrow A_s=A/\rho_{c0}^{1+\alpha}$ which allows us to recover the GCG perfect fluid case. 
It is useful to explore the cosmography parameters, mainly the deceleration and jerk parameters, to distinguish the viscous effects included in the GCG through the $\{q,j\}$-phase space, and also to compare VGCG with GCG and $\Lambda$CDM. 

The deceleration parameter can be estimated by $q(z)=-1 + (1+z)E(z)^{-1}( dE(z)/dz)$, then using \eqref{eq:E2} we obtain,
\begin{eqnarray} \label{eq:qz}
    &&q(z) = \frac{1}{2E(z)^2}\Big\lbrace-3(1-\sqrt{3}\xi_0)(1-B_s)\times\nonumber\\&&\Omega_{c0}(1+z)^{-3(1+\alpha)(1-\sqrt{3}\xi_0)}\Big[B_s+\nonumber\\&&(1-B_s)(1+z)^{-3(1+\alpha)(1-\sqrt{3}\xi_0)}\Big]^{-\alpha/(1+\alpha)} +\nonumber\\&&3\Omega_{b0}(z+1)^3+3\Omega_{r0}(z+1)^4   \Big\rbrace-1,
\end{eqnarray}
and the jerk parameter takes the form
\begin{equation} \label{eq:jz}
j(z) =  q(2q+1)+(1+z)\frac{dq}{dz},
\end{equation}
where $q$ is given by Eq. \eqref{eq:qz}. Finally, it is also useful to analyse the viscous effects through a test related to the well-known $Om$-diagnostic proposed in \cite{Sanhi:2008}, illustrated in a $E^2$ versus $(1+z)^3$ panel, that gives a clear evidence of whether the cosmological models behave as phantom  and/or quintessence and their deviations with respect to $\Lambda$CDM. In this context, $\Lambda$CDM is represented by a trajectory with constant slope and a upper (lower) trajectory for the other cosmological models over the $\Lambda$CDM one for $z>0$ suggests a quintessence (phanthom) behaviour. 

\section{Data and methodology} \label{sec:constraints}

A Bayesian Markov Chain Monte Carlo (MCMC) analysis is performed to constrain the phase-space parameter $\bf{\Theta}$ of the viscous generalized Chaplygin gas using OHD, SnIa, SLS, BAO, HIIG data and a joint analysis. Then, under the environment of the \texttt{emcee} Python package \cite{Foreman:2013}, we use the Gelman-Rubin criterion \cite{Gelman:1992} to achieve the convergence of the chains, and then we set 4000 chains with 250 steps each one to establish bounds on the model parameters. The phase-space priors are: 
 Gaussian for $\Omega_{b0}h^2=0.02242 \pm 0.00014$ \cite{Planck:2018} and
 flat distributions for the rest of the parameter in the ranges $h:[0.6,0.8]$, $B_s:[0,2]$, $\alpha:[0,2]$, and $\xi_0:[0,1]$. For the joint analysis, the figure-of-merit to be optimized is given by
\begin{equation}\label{eq:chi2}
    \chi_{\rm Joint}^2 = \chi_{\rm OHD}^2 + \chi_{\rm SnIa}^2 + \chi_{\rm SLS}^2 + \chi_{\rm BAO}^2 + \chi_{\rm HIIG}^2\,,
\end{equation}
where the $\chi^2$ sub indexes correspond to the names of the samples. The rest of the section is devoted to discuss the details of the samples used to constrain $\bf{\Theta}$ at background level. 

\subsection{Observational Hubble Parameter}

The observational Hubble parameter data (OHD) represents the most direct way to constrain the parameter space. The largest OHD sample is compiled by  \cite{Magana:2017nfs} and contains a total of 51 points covering the range $0<z<2.36$. From them, only 31 points are cosmological independent measurements obtained using differential age (DA) tools and the rest comes from BAO measurements. Because these measurements are considered to be uncorrelated, the chi square function can be expressed as
\begin{equation} \label{eq:chiOHD}
    \chi^2_{{\rm OHD}}=\sum_{i=1}^{51}\left(\frac{H_{th}(z_i)-H_{obs}(z_i)}{\sigma^i_{obs}}\right)^2,
\end{equation}
where $H_{th}(z_i)$ is the theoretical estimate using \eqref{eq:E2} and $H_{obs}(z_i)\pm \sigma_{obs}^i$ is the observational Hubble parameter with its uncertainty at the redshift $z_i$.

\subsection{Type Ia Supernovae }

Authors in Ref. \cite{Scolnic:2018} provide 1048 luminosity modulus measurements, known as Pantheon sample, from Type Ia Supernovae which cover a region $0.01<z<2.3$. Due to in this sample the measurements are correlated, it is convenient to build the chi square function as
\begin{equation}\label{eq:chi2SnIa}
    \chi_{\rm SnIa}^{2}=a +\log \left( \frac{e}{2\pi} \right)-\frac{b^{2}}{e},
\end{equation}
where
\begin{eqnarray}
    a &=& \Delta\boldsymbol{\tilde{\mu}}^{T}\cdot\mathbf{Cov_{P}^{-1}}\cdot\Delta\boldsymbol{\tilde{\mu}}, \nonumber\\
    b &=& \Delta\boldsymbol{\tilde{\mu}}^{T}\cdot\mathbf{Cov_{P}^{-1}}\cdot\Delta\mathbf{1}, \\
    e &=& \Delta\mathbf{1}^{T}\cdot\mathbf{Cov_{P}^{-1}}\cdot\Delta\mathbf{1}, \nonumber
\end{eqnarray}
and $\Delta\boldsymbol{\tilde{\mu}}$ is the vector of residuals between the theoretical distance modulus and the observed one, $\Delta\mathbf{1}=(1,1,\dots,1)^T$, $\mathbf{Cov_{P}}$ is the covariance matrix formed by adding the systematic and statistic uncertainties, i.e.   $\mathbf{Cov_{P}}=\mathbf{Cov_{P,sys}}+\mathbf{Cov_{P,stat}}$. The super-index $T$ on the above expressions denotes the transpose of the vectors.

The theoretical distance modulus is estimated by
\begin{equation}
    m_{th}=\mathcal{M}+5\log_{10}[d_L(z)/10\, pc],
\end{equation}
where $\mathcal{M}$ is a nuisance parameter which has been marginalized in \eqref{eq:chi2SnIa}. The dimensionless luminosity distance, denoted as $d_L(z)$, is computed by 
\begin{equation}\label{eq:dL}
    d_L(z)=(1+z)c\int_0^z\frac{dz^{\prime}}{H(z^{\prime})},
\end{equation}
where $c$ is the speed of light.

\subsection{Baryon Acoustic Oscillations}
Another way to establish a constraint of the model parameter is through the standard rules known as Baryon Acoustic Oscillations (BAO). These are primordial signatures produced by the interaction between baryons and photons in a hot plasma in the pre-recombination epoch. Authors in \cite{nunes:2020} collected 15 transversal BAO scale measurements which come from luminous red galaxies distributed in the redshift range $0.110<z<2.225$. As these points are considered uncorrelated, the chi square function is built as
\begin{equation}
\chi^2_{\rm BAO} = \sum_{i=1}^{15} \left( \frac{\theta_{\rm BAO}^i - \theta_{th}(z_i) }{\sigma_{\theta_{\rm BAO}^i}}\right)^2\,,
\end{equation}
where $\theta_{\rm BAO}^i \pm \sigma_{\theta_{\rm BAO}^i}$ is the BAO angular scale and its uncertainty at $68\%$ measured at $z_i$. On the other hand, the theoretical BAO angular scale, denoted as $\theta_{th}$, is estimated by
\begin{equation}
    \theta_{th}(z) = \frac{r_{drag}}{(1+z)D_A(z)}\,,
\end{equation}
where $D_A=d_L(z)/(1+z)^2$ is the angular diameter distance at $z$ and $d_L(z)$ was defined in \eqref{eq:dL}. The parameter
$r_{drag}$ is defined by the sound horizon at baryon drag epoch.
For this work, we set the $r_{drag}=147.21 \pm 0.23$ obtained by Planck collaboration \cite{Planck:2018}.

\subsection{Strong lensing systems}

A compilation of 204 strong lensing systems (SLS) in the redshift $0.0625<z_l<0.958$ for the lens and $0.196<z_s<3.595$ for the source is provided by \cite{Amante:2019xao}.
The chi square function for SLS takes the form
\begin{equation}\label{eq:chiSLS}
    \chi^2_{\rm SLS}=\sum_i^{204}\frac{[D^{th}(z_l,z_s)-D^{obs}(\theta_E,\sigma^2)]^2}{(\delta D^{obs})^2}\,,
\end{equation}
where the observable to confront is $D^{obs}=c^2\theta_E/4\pi\sigma^2$, where $\theta_E$ is the Einstein radius of the lens obtained by assuming the gravitational lens potential is modeled by a Singular Isothermal Sphere (SIS) defined by
\begin{equation}\label{eq:chiSLS}
    \theta_E=4\pi\frac{\sigma_{SIS}^2D_{ls}}{c^2D_s}\,.
\end{equation}
In the above expression, $\sigma_{SIS}$ is the velocity dispersion of the lens galaxy, $D_s$ is the angular diameter distance to the source, and $D_{ls}$ is the angular diameter distance from the lens to the source. It is interesting to remark that as this SLS data assumes a lens model for $\theta_E$ and $\sigma_{SIS}$ comes from spectroscopy, the sample is independent of $h$, and as consequence the parameter constraints do not depend on $h$.
The uncertainty of $D^{obs}$ is estimated by
\begin{equation}
    \delta D^{obs}=D^{obs}\left[\left(\frac{\delta\theta_E}{\theta_E}\right)^2+4\left(\frac{\delta\sigma}{\sigma}\right)^2\right]^{1/2}\,,
\end{equation}
where $\delta\theta_E$ and $\delta\sigma$ are the uncertainties of the Einstein radius and the observed velocity dispersion,  respectively.

The theoretical counterpart is estimated by the ratio
\begin{equation}
    D^{th}\equiv D_{ls}/D_s
\end{equation}
where $D_{ls}$ is the angular diameter distance from the lens to the source given by
\begin{equation}
         D_{ls}(z)=\frac{c}{1+z}\int_{z_l}^{z_s}\frac{dz^{\prime}}{H(z^{\prime})},
\end{equation}
and $D_s=d_L(z)/(1+z)^2$ is the angular diameter distance to the source which is obtained using \eqref{eq:dL}.

\subsection{HII Galaxies}

Authors \citep[][and references therein]{Chavez2012,Chavez2014,Terlevich2015,Chavez2016,GonzalezMoran2019} argued that the correlation between the measured luminosity $L$ and the inferred velocity dispersion $\sigma$ of the ionized gas (e.g. $H\beta$, $H\alpha$, $[OIII]$ emission lines) in extreme starburst galaxies (i.e. containing a population of O and/or B stars) may be used as a cosmological tracer to constrain cosmological model parameters.  A compilation of 153 HII galaxies (HIIG), containing apparent magnitude, emission line luminosity and velocity dispersion is provided by \cite{GonzalezMoran2019, Cao:2020jgu}. 
Then, the chi square function is estimated by
\begin{equation}\label{eq:chiHIIG}
    \chi^2_{{\rm HIIG}} = A - B^2/C\,,
\end{equation}
where
\begin{eqnarray}
    A &=& \sum_{i=1}^{153} \left( \frac{\mu_{th}(z_i)-\mu_{obs}^i}{\sigma_{\mu_{obs}^i}}\right)^2 \,,\\
    B &=& \sum_{i=1}^{153} \frac{\mu_{th}(z_i)-\mu_{obs}^i}{\sigma_{\mu_{obs}^i}} \,, \\
    C &=& \sum_{i=1}^{153} \frac{1}{(\sigma_{\mu_{obs}^i})^2}\,.
\end{eqnarray}
In the above expressions, $\mu_{obs}^i \pm \sigma_{obs}^i$ is the observed distance modulus with its uncertainty at redshift $z_i$. The theoretical estimate at the redshift $z$ is obtained by using \eqref{eq:dL} and
\begin{equation}
    \mu_{th}(z) = \mu_0 + 5 \log [\,d_L(z)\,] \,,
\end{equation}
where $\mu_0$ is a nuisance parameter which has been marginalized.

\section{Results} \label{sec:Results}

\begin{figure}
\centering
\par\smallskip
{\includegraphics[width=0.45\textwidth]{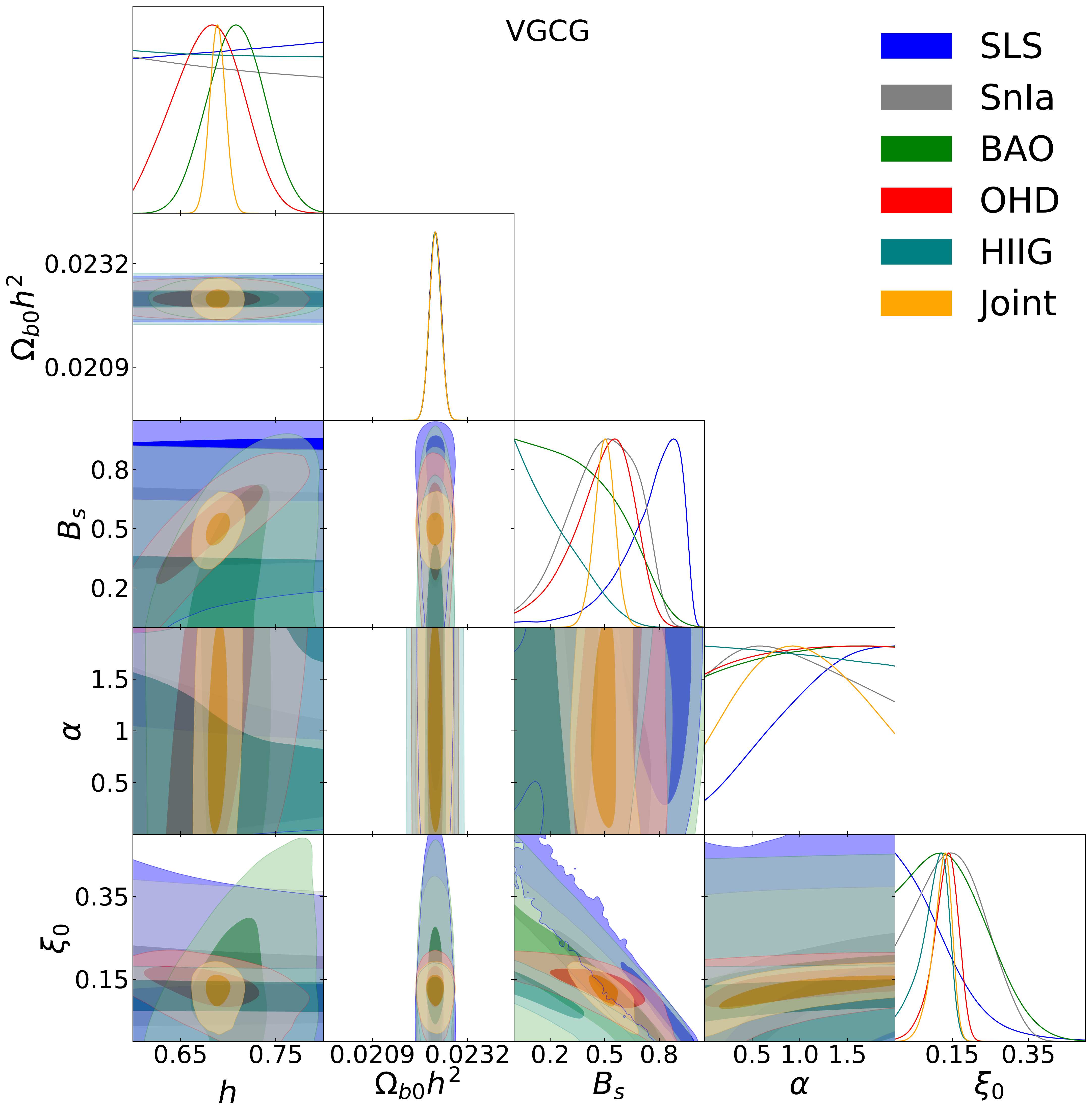}}\\
{\includegraphics[width=0.45\textwidth]{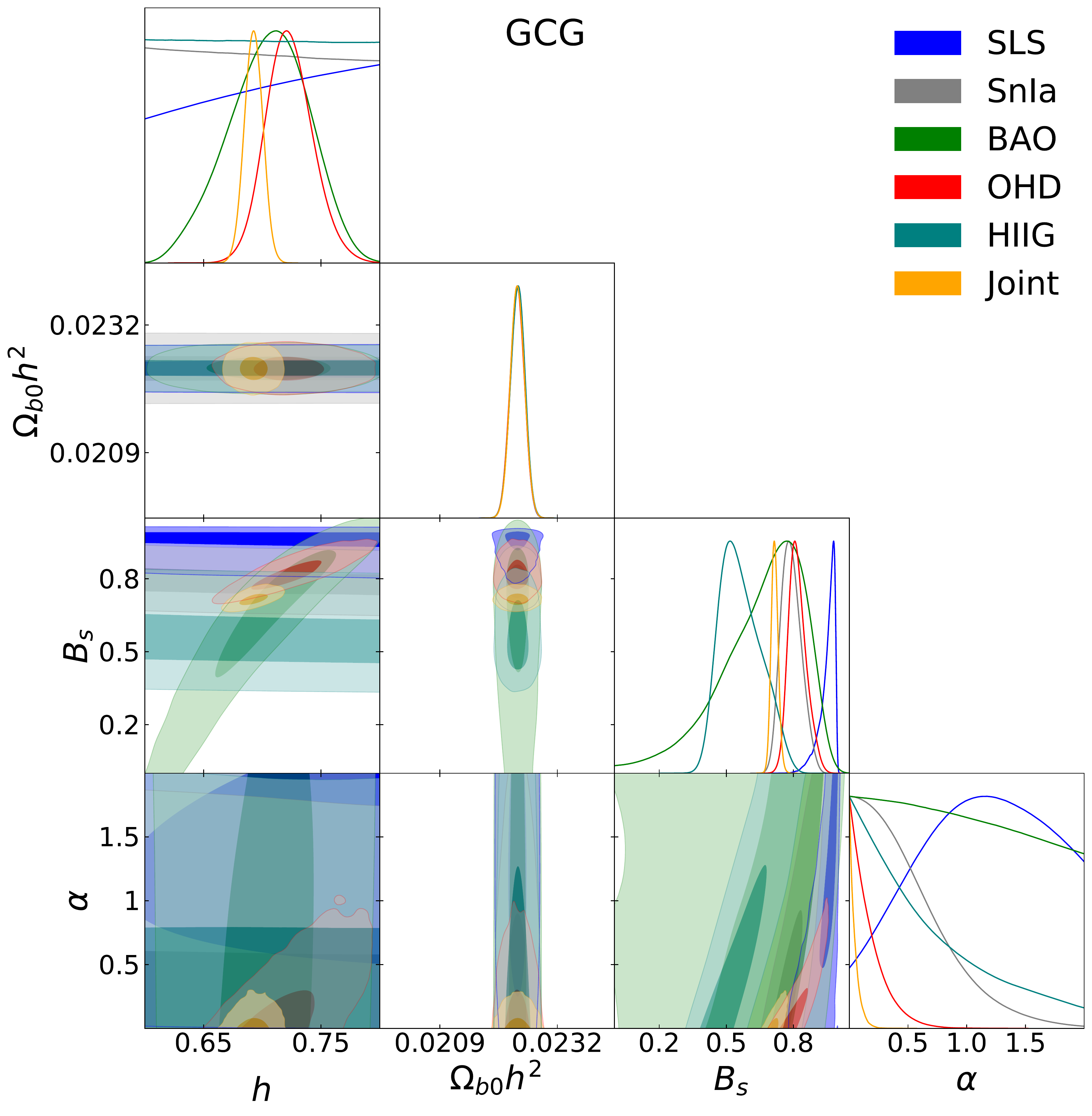}}
\caption{1D marginalized posterior distributions and the 2D distributions at $1\sigma$ (dark region) and $3\sigma$ (light region) CL for the parameters space $\Theta$.}
\label{fig:contours}
\end{figure}

Table \ref{tab:par} presents the cosmological constraints of VGCG and GCG for OHD, SnIa, SLS, BAO, HIIG samples and the joint analysis respectively. Each best-fit parameter value includes its uncertainty at $68\%$ confidence level (CL) and is consistent, within $1\sigma$, with those reported in the literature \cite{Li-Xu:2013}. 
Figure \ref{fig:contours} shows the 1D marginalized posterior distributions for each data and joint analysis and also the 2D phase space distribution at $68\%\,(1\sigma)$, $99.7\%\,(3\sigma)$ CL. According to the $\chi^2$ value, both models are in good agreement with the data\footnote{Notice that, in the case of the HIIG sample, $\chi^2/dof$ is much larger than 1 for both models. This might indicate an underestimation of the observational errors.}. However, as the VGCG includes an extra parameter over GCG, it is better to confront them statistically using the corrected Akaike information criterion (AICc) \cite{AIC:1974, Sugiura:1978, AICc:1989} and the Bayesian information criterion (BIC) \cite{schwarz1978} defined as ${\rm AICc}= \chi^2_{min}+2k +(2k^2+2k)/(N-k-1)$ 
and ${\rm BIC}=\chi^2_{min}+k\log(N)$ respectively, where $k$ is the number of free parameters and $N$ is the number of data points. The criteria establishes that the model with lower values of AICc and BIC is preferred by data. For AICc, if a difference between a given model and the best one, $\Delta\rm{AICc}$, 
is $\Delta \rm{AICc}<4$, both models are supported by the data equally. 
If $4<\Delta\rm{AICc}<10$ the data still support the given model but less than the preferred one.
A value of $\Delta \rm{AICc}>10$ indicates that the data does not support the given model.
For BIC, the difference between a candidate model and the best model $\Delta \rm{BIC}$ is interpreted as the evidence against the candidate model being the best model. 
A yield of $\Delta \rm{BIC}<2$ indicates there is no evidence against the candidate model.
A value in the range $2<\Delta\rm{BIC}<6$ suggests  that there is modest evidence against the candidate model. 
A strong evidence against the candidate model is given when $6<\Delta \rm{BIC}<10$, and 
a stronger evidence against is if $\Delta \rm{BIC}>10$.
In summary, based on our $\Delta$AICc results given in Table \ref{tab:par}, the data show a similar preference for both models, VGCG and GCG. In contrast, $\Delta$BIC results give a strong evidence against VGCG for SLS and SnIa data and no evidence against for the rest of the samples, including the combined sample (Joint).
These results are expected considering that BIC penalizes free parameters more strongly than AICc and that the penalization function is proportional to the number of points in the dataset.

\begin{table*}
\caption{Best fitting values of VGCG and GCG for each dataset. The uncertainties are at $1\sigma$ CL.}
\resizebox{0.95\textwidth}{!}{
\begin{tabular}{|ccccccc|}
\hline
Data set & $\chi^{2}_{min}/$dof & $h$ & $\Omega_{b0}h^2$ &$B_s$ & $\alpha$ & $\xi_0$ \\
\hline
\multicolumn{7}{|c|}{VGCG}\\
\hline
SLS & $570.5/199$ & $0.74^{+0.18}_{-0.20}$  & $0.02242^{+0.00014}_{-0.00014}$  & $0.79^{+0.12}_{-0.21}$  & $1.29^{+0.49}_{-0.66}$  & $0.09^{+0.10}_{-0.06}$  \\ [.8ex] 
SnIa & $1036.4/1043$ & $0.70^{+0.20}_{-0.18}$  & $0.02242^{+0.00014}_{-0.00014}$  & $0.50^{+0.18}_{-0.21}$  & $0.93^{+0.69}_{-0.61}$  & $0.15^{+0.09}_{-0.09}$  \\ [.8ex] 
BAO & $14.3/10$ & $0.71^{+0.03}_{-0.03}$  & $0.02242^{+0.00014}_{-0.00014}$  & $0.33^{+0.27}_{-0.23}$  & $1.03^{+0.66}_{-0.68}$  & $0.14^{+0.11}_{-0.09}$  \\ [.8ex] 
OHD & $19.9/46$ & $0.68^{+0.04}_{-0.04}$  & $0.02242^{+0.00014}_{-0.00014}$  & $0.50^{+0.14}_{-0.20}$  & $1.03^{+0.66}_{-0.68}$  & $0.13^{+0.03}_{-0.04}$  \\ [.8ex] 
HIIG & $2202.6/148$ & $0.72^{+0.19}_{-0.19}$  & $0.02242^{+0.00014}_{-0.00014}$  & $0.18^{+0.21}_{-0.13}$  & $0.97^{+0.69}_{-0.67}$  & $0.11^{+0.03}_{-0.05}$  \\ [.8ex] 
Joint & $3942.9/1466$ & $0.69^{+0.01}_{-0.01}$  & $0.02242^{+0.00014}_{-0.00014}$  & $0.50^{+0.05}_{-0.06}$  & $0.99^{+0.61}_{-0.58}$  & $0.13^{+0.02}_{-0.03}$  \\ [.8ex] 
\hline
\multicolumn{7}{|c|}{GCG}\\
\hline
SLS & $568.8/200$ & $0.78^{+0.16}_{-0.19}$  & $0.02242^{+0.00014}_{-0.00014}$  & $0.96^{+0.02}_{-0.04}$  & $1.13^{+0.56}_{-0.59}$ &--  \\ [.8ex] 
SnIa & $1036.3/1044$ & $0.71^{+0.20}_{-0.18}$  & $0.02242^{+0.00014}_{-0.00014}$  & $0.79^{+0.05}_{-0.04}$  & $0.41^{+0.44}_{-0.28}$&--  \\ [.8ex] 
BAO & $13.0/11$ & $0.71^{+0.03}_{-0.04}$  & $0.02242^{+0.00014}_{-0.00014}$  & $0.69^{+0.15}_{-0.22}$  & $0.93^{+0.71}_{-0.64}$&--  \\ [.8ex] 
OHD & $23.0/47$ & $0.72^{+0.02}_{-0.02}$  & $0.02242^{+0.00014}_{-0.00014}$  & $0.81^{+0.05}_{-0.03}$  & $0.12^{+0.18}_{-0.09}$ &-- \\ [.8ex] 
HIIG & $2204.5/149$ & $0.72^{+0.19}_{-0.19}$  & $0.02243^{+0.00014}_{-0.00014}$  & $0.55^{+0.11}_{-0.08}$  & $0.49^{+0.71}_{-0.36}$&--  \\ [.8ex] 
Joint & $3952.7/1467$ & $0.69^{+0.01}_{-0.01}$  & $0.02242^{+0.00014}_{-0.00014}$  & $0.72^{+0.02}_{-0.01}$  & $0.03^{+0.05}_{-0.02}$&--  \\ [.8ex] 
\hline
\end{tabular}}
\label{tab:par}
\end{table*}

\begin{table*}
\caption{AICc and BIC criteria for VGCG and GCG. The difference between models are $\Delta \rm{AICc}=\rm{AICc}^{VGCG}-\rm{AICc}^{GCG}$ and $\Delta \rm{BIC}=\rm{BIC}^{VGCG}-\rm{BIC}^{GCG}$.}
\begin{tabular}{|ccccccc|}
\hline
Data set & AICc$^{\rm{VGCG}}$ & AICc$^{\rm{GCG}}$ & $|\Delta\rm{AICc}|$ & BIC$^{\rm{VGCG}}$ & BIC$^{\rm{GCG}}$ & $|\Delta\rm{BIC}|$ \\
\hline
SLS   & $580.8$ 	 & $577.0$  	 & $3.8$ 	 & $597.1$  	 & $590.1$ 	     & $7.0$ \\ [.8ex]
SnIa  & $1046.5$ 	 & $1044.3$ 	 & $2.1$ 	 & $1071.2$ 	 & $1064.1$ 	 & $7.1$ \\ [.8ex]
BAO   & $31.0$ 	     & $25.0$ 	     & $6.0$ 	 & $27.8$ 	     & $23.8$   	 & $4.0$ \\ [.8ex]
OHD   & $31.2$ 	     & $31.9$   	 & $0.6$ 	 & $39.6$ 	     & $38.7$ 	     & $0.8$ \\ [.8ex]
HIIG  & $2213.0$ 	 & $2212.8$ 	 & $0.2$ 	 & $2227.8$ 	 & $2224.6$ 	 & $3.1$ \\ [.8ex]
Joint & $3952.9$ 	 & $3960.7$ 	 & $7.8$ 	 & $3979.4$ 	 & $3981.9$ 	 & $2.5$ \\ [.8ex]
\hline
\end{tabular}
\label{tab:AIC}
\end{table*}

\begin{figure}
\centering
\par\smallskip
{\includegraphics[width=0.45\textwidth]{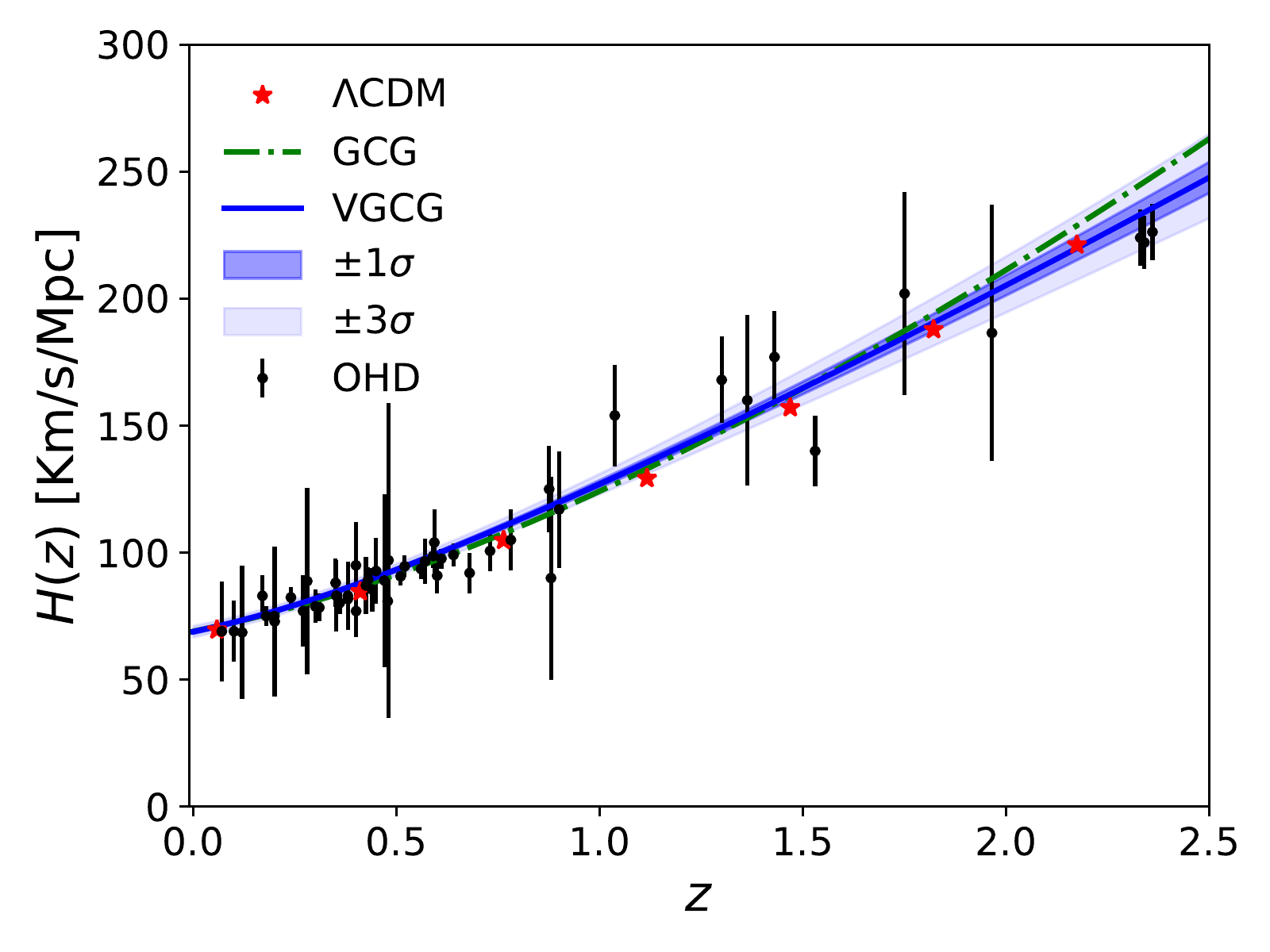}}
\caption{Best fitting curve (joint analysis) over OHD points for the GCG, VGCG  and its comparison with $\Lambda$CDM model. The bands correspond to the uncertainty of VGCG at $1\sigma$ (darker band) and $3\sigma$ (lighter band).}
\label{Figure:fit_OHD}
\end{figure}

\subsection{Cosmography parameters}

Based on the joint analysis, the parameters of deceleration ($q$) and jerk ($j$) at $z=0$ are estimated, obtaining respectively values of
$q_0 = -0.523^{+0.021}_{-0.024}$,
$j_0 = 1.029^{+0.042}_{-0.021}$
for the GCG and
$q_0 = -0.514^{+0.063}_{-0.064}$,
$j_0 = 1.319^{+0.283}_{-0.295}$
for VGCG where the uncertainties are at $1\sigma$. These $q_0$ values are in agreement with the $\Lambda$CDM value \cite{Herrera:2020} within $1.6\sigma$ and $1.1\sigma$ respectively. 
Regarding the jerk parameter, our results are consistent within $1\sigma$ with respect to the expected $\Lambda$CDM value. 
Additionally, it is interesting to compare our $q_0$ ($j_0$) values of VGCG with other viscous models.
For instance, we find a good agreement with respect to models which consider a constant viscosity within $1.5\sigma$ ($0.6\sigma$) and polynomial viscosity within $1\sigma$ ($1.2\sigma$) (see Fig. 4 of \cite{Herrera:2020}). It is worth mention that although the VGCG model differs from the mentioned models in the EoS ($w=-1$) and the viscosity coefficient (constant and polynomial), we find consistent results in the current values of $q$ and $j$. 

\begin{figure}
\centering
\par\smallskip
{\includegraphics[width=0.45\textwidth]{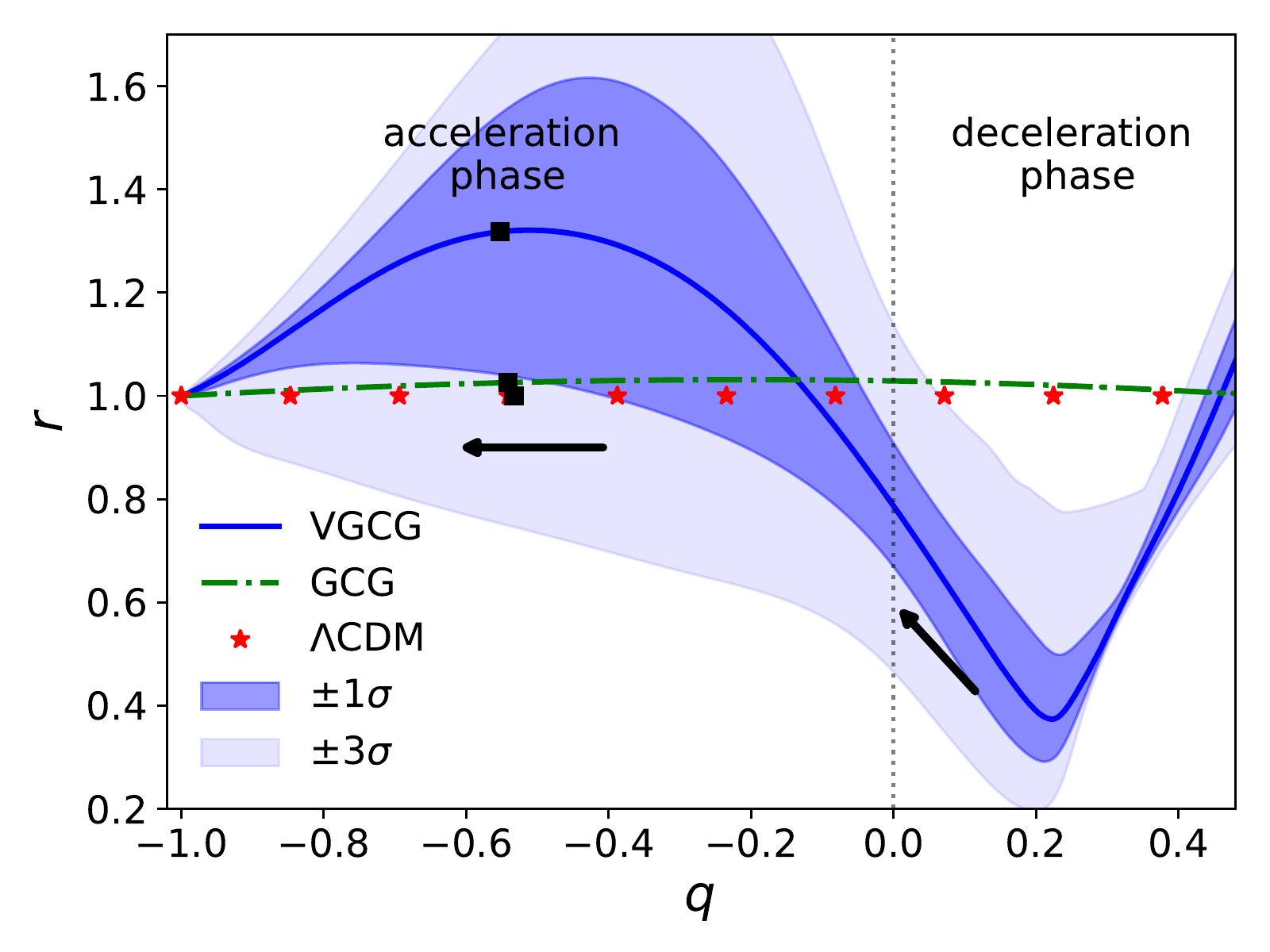}}
\caption{\{q,r\}-space reconstruction on $z$ for VGCG, GCG, and $\Lambda$CDM. Arrows represent the direction of the universe evolution, and black squares are the current states. The bands correspond to the uncertainty of VGCG at $1\sigma$ (darker band) and $3\sigma$ (lighter band).}
\label{Figure:qvsr}
\end{figure}

An alternative way to find differences between cosmological models is through the \{$q,r$\}-panel where $r=j$\footnote{Typically this space is shown by using the notation $r$ instead of $j$.}. Figure \ref{Figure:qvsr} shows such \{$q,r$\}-space reconstruction for both models and $\Lambda$CDM in the range $-1<z<150$. The arrows represent the direction of the universe evolution, being the evolution from a deceleration phase to an accelerated epoch. The black square markers over the trajectories represent the current states and the bands around the VGCG trajectory are its uncertainties up to $3\sigma$. It is interesting to observe that for both models, their states {$q,r$} converge to $\Lambda$CDM state in the future ($z\to -1$) in contrast of the observed behaviour by considering other viscous models as those presented in \cite{Herrera:2020}. Additionally, we observe the path of GCG is close to the one by $\Lambda$CDM, while for the VGCG we find a deviation of more than $3\sigma$ in the deceleration epochs from $\Lambda$CDM model.

Additionally, high order parameters known as snap ($s$), and lent ($l$) at current time are calculated, we obtain yields
$s_0 = -0.494^{+0.062}_{-0.065}$,
$l_0 = 3.526^{+0.190}_{-0.171}$
and 
$s_0 = -0.482^{+0.225}_{-0.611}$,
$l_0 = -0.110^{+1.960}_{-3.701}$
for GCG and VGCG respectively. We obtain $s_0$ values for VGCG (GCG) deviated up to $1.6\sigma$ ($2.9\sigma$) and $1.6\sigma$ ($3.1\sigma$) from those reported in \cite{Herrera:2020}.

\subsection{Near equilibrium condition}

\begin{figure}
\centering
\par\smallskip
{\includegraphics[width=0.45\textwidth]{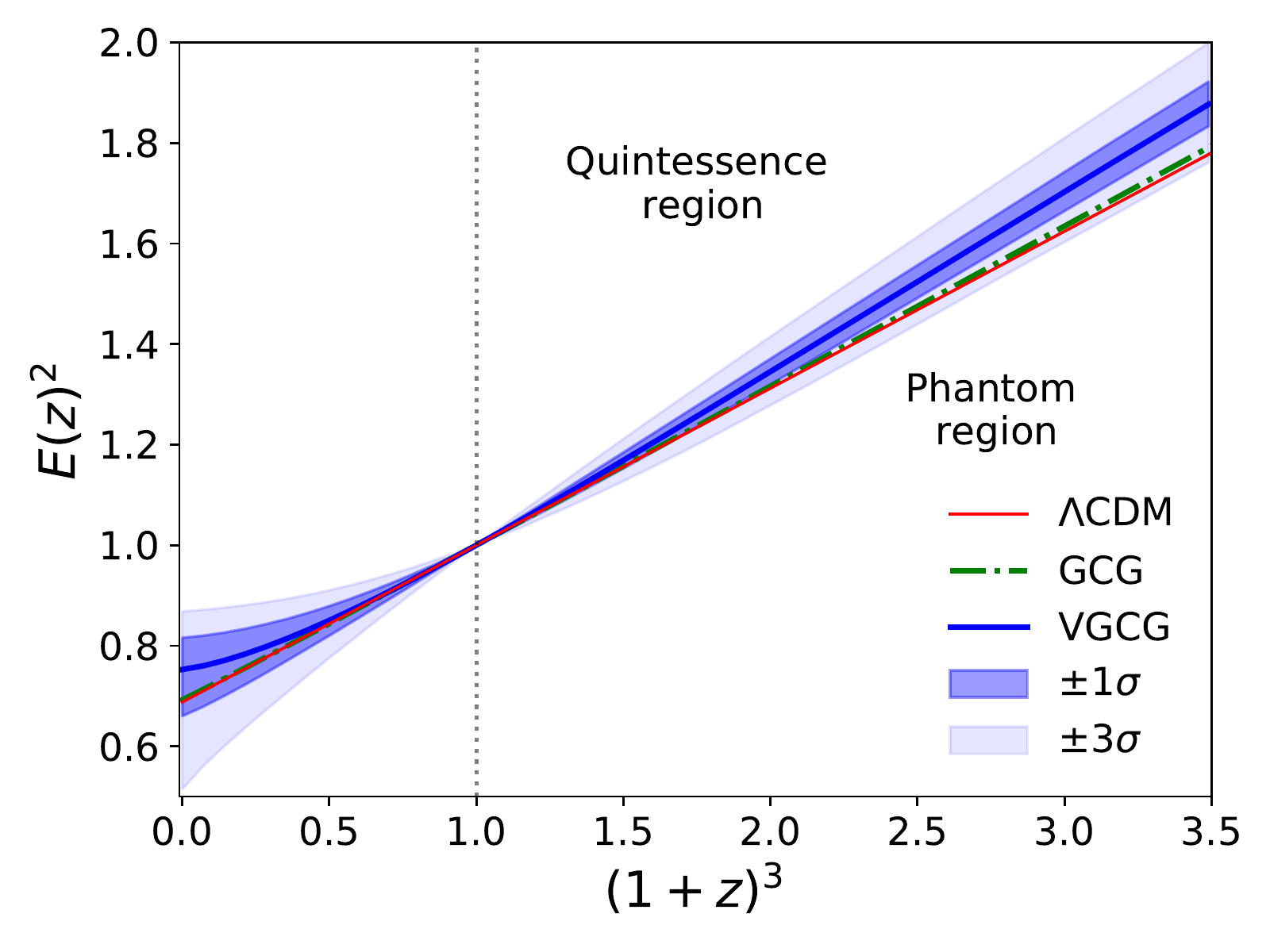}}
\caption{$E(z)^2$ reconstruction over $(1+z)^3$ for VGCG, GCG, and $\Lambda$CDM. The bands correspond to the uncertainty of VGCG at $1\sigma$ (darker band) and $3\sigma$ (lighter band).}
\label{Figure:Omdiag}
\end{figure}

Figure \ref{Figure:Omdiag} displays $E(z)^2$ with respect to $(1+z)^3$ for VGCG, GCG and $\Lambda$CDM. This panel shows that VGCG and GCG behave as quintessence in the past ($0<z<2.5$) and as phantom in the future ($-1<z<0$). Furthermore, we observe agreement up to $3\sigma$ between VGCG and GCG, and a deviation of more than $3\sigma$ between VGCG and $\Lambda$CDM in $z\gtrsim 2.3$.

\begin{figure}
\centering
\par\smallskip
{\includegraphics[width=0.45\textwidth]{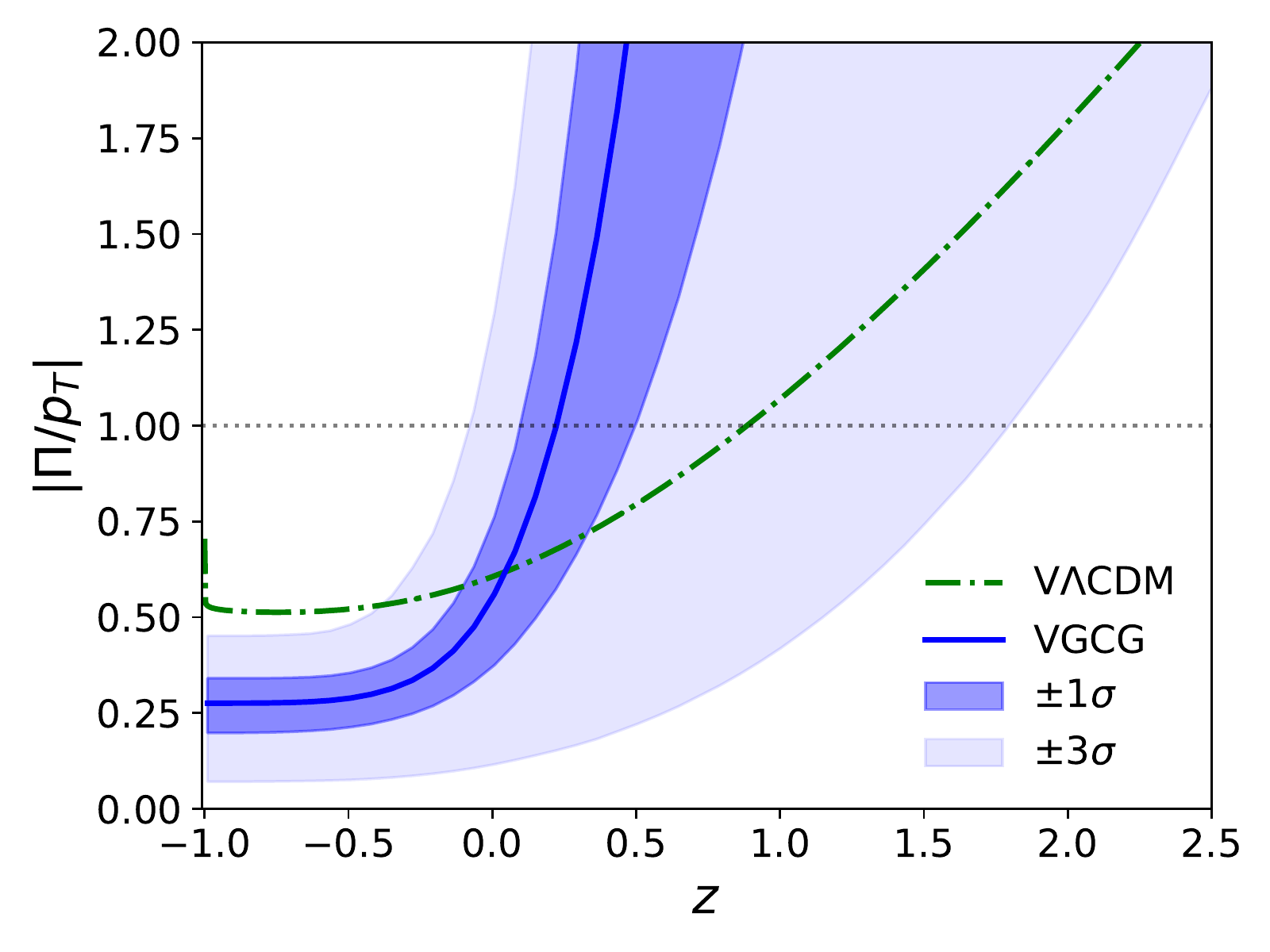}}
\caption{$|\Pi/p_T|$ reconstruction over $z$ for VGCG. The blue bands correspond to the uncertainty of VGCG at $1\sigma$ (darker band) and $3\sigma$ (lighter band). The dot-dashed (green) line represents the case of $\Lambda$CDM in presence of viscosity.}
\label{Figure:Pip}
\end{figure}

Because the VGCG model produces an accelerated expansion stage using two mechanism (the viscous pressure and the EoS), this model does not need to satisfy the near equilibrium condition required in the Eckart formalism. In this context, a natural question is how far the VGCG is from this condition given by  $ \left | \Pi/p_T \right | \ll 1$,
where $p_T$ is the total pressure of the fluids. Figure \ref{Figure:Pip} shows the behavior of $|\Pi/p_T|$ for VGCG with its uncertainties up to $3\sigma$ (blue bands) and the case of $\Lambda$CDM in presence of viscosity (V$\Lambda$CDM) when $\xi\sim \rho^{1/2}$ \cite{Normann:2016}. Although, we find a good agreement between both models, within $3\sigma$ in the region $-0.5<z<2.5$, we observe that VGCG (V$\Lambda$CDM) presents a redshift region $z\gtrsim 0.2$ ($z\gtrsim 0.9$) 
with 
$|\Pi/p_T|>1$ which are values lower than those found in \cite{Herrera:2020}, who find a redshift region $z\gtrsim 4$ ($z\gtrsim 13$) by considering a polynomial (constant) bulk viscosity using OHD+SnIa+SLS data. 

\section{Conclusions and Outlooks} \label{sec:summary}

This paper was dedicated to revisit a hybrid model that contains a non-perfect Chaplygin gas, called VGCG, to produce the accelerated expansion stage. VGCG was studied mainly in two parts at the background level. The first part was devoted to perform a Bayesian statistics analysis to confront the model with the most recent cosmological data (OHD, SnIa, BAO, SLS, HII Galaxies) and give updated values of the VGCG parameters. We found consistent constraints of the phase space parameters within $1\sigma$. Based on our Joint analysis results shown in Table \ref{tab:par}, the second part was dedicated to explore the evolution of the cosmographic parameters and the near equilibrium condition required in the Eckart formalism. In particular, we analysed the $\{q,r\}$-space which is useful to find differences between cosmological models. Thus, we observed a deviation of VGCG with respect to the $\Lambda$CDM and GCG of more than $3\sigma$ CL in the deceleration phase (see Figure \ref{Figure:qvsr}). Additionally, we reported the current cosmographic values
$q_0 = -0.514^{+0.063}_{-0.064}$,
$j_0 = 1.319^{+0.283}_{-0.295}$,
$s_0 = -0.482^{+0.225}_{-0.611}$ and
$l_0 = -0.110^{+1.960}_{-3.701}$ for the VGCG model corresponding to deceleration, jerk, snap, and lent parameters respectively. We obtained consistent values with those values reported in \cite{Herrera:2020} for alternative viscous models.
On the other hand, we found a quintessence behaviour for VGCG and GCG models in the past $z>0$ and a phantom one for the future as shown in Figure \ref{Figure:Omdiag}, with the difference that GCG presents divergence at $z=-1$. 
On the other hand, we reconstructed the $\vert\Pi/p_T\vert$ to explore how far the VGCG is from the near equilibrium condition. We found consistent results up to $3\sigma$ with the behaviour by considering a $\Lambda$CDM model with viscosity presented in \cite{Normann:2016}. In this context, we observed that VGCG presents a tighter phase around current epochs which $\vert\Pi/p_T\vert<1$ with respect to those regions for the viscous $\Lambda$CDM reported in \cite{Normann:2016,Herrera:2020}.

Finally, based on our results presented in Table \ref{tab:par}, it is noteworthy that the VGCG suggests a central value on $h$ between the value obtained by Planck \cite{Planck:2018} and the value by Riess et. al. \cite{Riess:2019}, contributing to diminish the tension between these measurements. In fact, we remark that although we are using data based on local measurements such as OHD that favor the  local value (Riess value), we obtain a central value on $H_0$ close to the Planck one. Hence, VGCG is a promising candidate to lessen the discrepancy generated in $H_0$. Further studies of these models is needed.

\bigskip

\acknowledgments{
We thank the anonymous referee for thoughtful remarks and suggestions. AHA thanks support from SNI-M\'exico. MAG-A acknowledges support from SNI-M\'exico, CONACyT research fellow, COZCyT and Instituto Avanzado de Cosmolog\'ia (IAC) collaborations. VM acknowledges the support of Centro de Astrof\'{\i}sica de Valpara\'{\i}so (CAV). MAG-A and VM acknowledge CONICYT REDES (190147). This work received support from Luis Aguilar, Alejandro de Le\'on, Carlos Flores, and Jair Garc\'ia of the Laboratorio Nacional de Visualizaci\'on Cient\'ifica Avanzada}

\bibliography{main}

\begin{thebibliography}{75}%
\makeatletter
\providecommand \@ifxundefined [1]{%
 \@ifx{#1\undefined}
}%
\providecommand \@ifnum [1]{%
 \ifnum #1\expandafter \@firstoftwo
 \else \expandafter \@secondoftwo
 \fi
}%
\providecommand \@ifx [1]{%
 \ifx #1\expandafter \@firstoftwo
 \else \expandafter \@secondoftwo
 \fi
}%
\providecommand \natexlab [1]{#1}%
\providecommand \enquote  [1]{``#1''}%
\providecommand \bibnamefont  [1]{#1}%
\providecommand \bibfnamefont [1]{#1}%
\providecommand \citenamefont [1]{#1}%
\providecommand \href@noop [0]{\@secondoftwo}%
\providecommand \href [0]{\begingroup \@sanitize@url \@href}%
\providecommand \@href[1]{\@@startlink{#1}\@@href}%
\providecommand \@@href[1]{\endgroup#1\@@endlink}%
\providecommand \@sanitize@url [0]{\catcode `\\12\catcode `\$12\catcode
  `\&12\catcode `\#12\catcode `\^12\catcode `\_12\catcode `\%12\relax}%
\providecommand \@@startlink[1]{}%
\providecommand \@@endlink[0]{}%
\providecommand \url  [0]{\begingroup\@sanitize@url \@url }%
\providecommand \@url [1]{\endgroup\@href {#1}{\urlprefix }}%
\providecommand \urlprefix  [0]{URL }%
\providecommand \Eprint [0]{\href }%
\providecommand \doibase [0]{http://dx.doi.org/}%
\providecommand \selectlanguage [0]{\@gobble}%
\providecommand \bibinfo  [0]{\@secondoftwo}%
\providecommand \bibfield  [0]{\@secondoftwo}%
\providecommand \translation [1]{[#1]}%
\providecommand \BibitemOpen [0]{}%
\providecommand \bibitemStop [0]{}%
\providecommand \bibitemNoStop [0]{.\EOS\space}%
\providecommand \EOS [0]{\spacefactor3000\relax}%
\providecommand \BibitemShut  [1]{\csname bibitem#1\endcsname}%
\let\auto@bib@innerbib\@empty
\bibitem [{\citenamefont {Aghanim}\ \emph {et~al.}(2018)\citenamefont {Aghanim}
  \emph {et~al.}}]{Planck:2018}%
  \BibitemOpen
  \bibfield  {author} {\bibinfo {author} {\bibfnamefont {N.}~\bibnamefont
  {Aghanim}} \emph {et~al.} (\bibinfo {collaboration} {Planck}),\ }\href@noop
  {} {\  (\bibinfo {year} {2018})},\ \Eprint {http://arxiv.org/abs/1807.06209}
  {arXiv:1807.06209 [astro-ph.CO]} \BibitemShut {NoStop}%
\bibitem [{\citenamefont {Riess}\ \emph {et~al.}(1998)\citenamefont {Riess},
  \citenamefont {Filippenko}, \citenamefont {Challis}, \citenamefont
  {Clocchiatti}, \citenamefont {Diercks} \emph {et~al.}}]{Riess:1998}%
  \BibitemOpen
  \bibfield  {author} {\bibinfo {author} {\bibfnamefont {A.~G.}\ \bibnamefont
  {Riess}}, \bibinfo {author} {\bibfnamefont {A.~V.}\ \bibnamefont
  {Filippenko}}, \bibinfo {author} {\bibfnamefont {P.}~\bibnamefont {Challis}},
  \bibinfo {author} {\bibfnamefont {A.}~\bibnamefont {Clocchiatti}}, \bibinfo
  {author} {\bibfnamefont {A.}~\bibnamefont {Diercks}},  \emph {et~al.},\
  }\href {http://stacks.iop.org/1538-3881/116/i=3/a=1009} {\bibfield  {journal}
  {\bibinfo  {journal} {The Astronomical Journal}\ }\textbf {\bibinfo {volume}
  {116}},\ \bibinfo {pages} {1009} (\bibinfo {year} {1998})}\BibitemShut
  {NoStop}%
\bibitem [{\citenamefont {Perlmutter}\ \emph {et~al.}(1999)\citenamefont
  {Perlmutter}, \citenamefont {Aldering}, \citenamefont {Goldhaber},
  \citenamefont {Knop}, \citenamefont {Nugent}, \citenamefont {others},\ and\
  \citenamefont {Project}}]{Perlmutter:1999}%
  \BibitemOpen
  \bibfield  {author} {\bibinfo {author} {\bibfnamefont {S.}~\bibnamefont
  {Perlmutter}}, \bibinfo {author} {\bibfnamefont {G.}~\bibnamefont
  {Aldering}}, \bibinfo {author} {\bibfnamefont {G.}~\bibnamefont {Goldhaber}},
  \bibinfo {author} {\bibfnamefont {R.~A.}\ \bibnamefont {Knop}}, \bibinfo
  {author} {\bibfnamefont {P.}~\bibnamefont {Nugent}}, \bibinfo {author}
  {\bibnamefont {others}}, \ and\ \bibinfo {author} {\bibfnamefont {T.~S.~C.}\
  \bibnamefont {Project}},\ }\href
  {http://stacks.iop.org/0004-637X/517/i=2/a=565} {\bibfield  {journal}
  {\bibinfo  {journal} {The Astrophysical Journal}\ }\textbf {\bibinfo {volume}
  {517}},\ \bibinfo {pages} {565} (\bibinfo {year} {1999})}\BibitemShut
  {NoStop}%
\bibitem [{\citenamefont {Scolnic}\ \emph {et~al.}(2018)\citenamefont {Scolnic}
  \emph {et~al.}}]{Scolnic:2018}%
  \BibitemOpen
  \bibfield  {author} {\bibinfo {author} {\bibfnamefont {D.~M.}\ \bibnamefont
  {Scolnic}} \emph {et~al.},\ }\href {\doibase 10.3847/1538-4357/aab9bb}
  {\bibfield  {journal} {\bibinfo  {journal} {Astrophys. J.}\ }\textbf
  {\bibinfo {volume} {859}},\ \bibinfo {pages} {101} (\bibinfo {year}
  {2018})},\ \Eprint {http://arxiv.org/abs/1710.00845} {arXiv:1710.00845
  [astro-ph.CO]} \BibitemShut {NoStop}%
\bibitem [{\citenamefont {Nadathur}\ \emph {et~al.}(2020)\citenamefont
  {Nadathur}, \citenamefont {Percival}, \citenamefont {Beutler},\ and\
  \citenamefont {Winther}}]{LargeScale}%
  \BibitemOpen
  \bibfield  {author} {\bibinfo {author} {\bibfnamefont {S.}~\bibnamefont
  {Nadathur}}, \bibinfo {author} {\bibfnamefont {W.~J.}\ \bibnamefont
  {Percival}}, \bibinfo {author} {\bibfnamefont {F.}~\bibnamefont {Beutler}}, \
  and\ \bibinfo {author} {\bibfnamefont {H.~A.}\ \bibnamefont {Winther}},\
  }\href {\doibase 10.1103/PhysRevLett.124.221301} {\bibfield  {journal}
  {\bibinfo  {journal} {Phys. Rev. Lett.}\ }\textbf {\bibinfo {volume} {124}},\
  \bibinfo {pages} {221301} (\bibinfo {year} {2020})}\BibitemShut {NoStop}%
\bibitem [{\citenamefont {Carroll}(2001)}]{Carroll:2000}%
  \BibitemOpen
  \bibfield  {author} {\bibinfo {author} {\bibfnamefont {S.~M.}\ \bibnamefont
  {Carroll}},\ }\href {\doibase 10.12942/lrr-2001-1} {\bibfield  {journal}
  {\bibinfo  {journal} {Living Rev. Rel.}\ }\textbf {\bibinfo {volume} {4}},\
  \bibinfo {pages} {1} (\bibinfo {year} {2001})},\ \Eprint
  {http://arxiv.org/abs/astro-ph/0004075} {arXiv:astro-ph/0004075 [astro-ph]}
  \BibitemShut {NoStop}%
\bibitem [{\citenamefont {Zeldovich}(1968)}]{Zeldovich}%
  \BibitemOpen
  \bibfield  {author} {\bibinfo {author} {\bibfnamefont {Y.~B.}\ \bibnamefont
  {Zeldovich}},\ }\href@noop {} {\bibfield  {journal} {\bibinfo  {journal}
  {Soviet Physics Uspekhi}\ }\textbf {\bibinfo {volume} {11}} (\bibinfo {year}
  {1968})}\BibitemShut {NoStop}%
\bibitem [{\citenamefont {Weinberg}(1989)}]{Weinberg}%
  \BibitemOpen
  \bibfield  {author} {\bibinfo {author} {\bibfnamefont {S.}~\bibnamefont
  {Weinberg}},\ }\href@noop {} {\bibfield  {journal} {\bibinfo  {journal}
  {Reviews of Modern Physics}\ }\textbf {\bibinfo {volume} {61}} (\bibinfo
  {year} {1989})}\BibitemShut {NoStop}%
\bibitem [{\citenamefont {De~Felice}\ and\ \citenamefont
  {Tsujikawa}(2010)}]{DeFelice:2010aj}%
  \BibitemOpen
  \bibfield  {author} {\bibinfo {author} {\bibfnamefont {A.}~\bibnamefont
  {De~Felice}}\ and\ \bibinfo {author} {\bibfnamefont {S.}~\bibnamefont
  {Tsujikawa}},\ }\href {\doibase 10.12942/lrr-2010-3} {\bibfield  {journal}
  {\bibinfo  {journal} {Living Rev. Rel.}\ }\textbf {\bibinfo {volume} {13}},\
  \bibinfo {pages} {3} (\bibinfo {year} {2010})},\ \Eprint
  {http://arxiv.org/abs/1002.4928} {arXiv:1002.4928 [gr-qc]} \BibitemShut
  {NoStop}%
\bibitem [{\citenamefont {Jaime}\ \emph {et~al.}(2011)\citenamefont {Jaime},
  \citenamefont {Patino},\ and\ \citenamefont {Salgado}}]{Jaime:2010kn}%
  \BibitemOpen
  \bibfield  {author} {\bibinfo {author} {\bibfnamefont {L.~G.}\ \bibnamefont
  {Jaime}}, \bibinfo {author} {\bibfnamefont {L.}~\bibnamefont {Patino}}, \
  and\ \bibinfo {author} {\bibfnamefont {M.}~\bibnamefont {Salgado}},\ }\href
  {\doibase 10.1103/PhysRevD.83.024039} {\bibfield  {journal} {\bibinfo
  {journal} {Phys. Rev.}\ }\textbf {\bibinfo {volume} {D83}},\ \bibinfo {pages}
  {024039} (\bibinfo {year} {2011})},\ \Eprint {http://arxiv.org/abs/1006.5747}
  {arXiv:1006.5747 [gr-qc]} \BibitemShut {NoStop}%
\bibitem [{\citenamefont {Jaime}\ \emph {et~al.}(2014)\citenamefont {Jaime},
  \citenamefont {Salgado},\ and\ \citenamefont {Patino}}]{Jaime:2012nj}%
  \BibitemOpen
  \bibfield  {author} {\bibinfo {author} {\bibfnamefont {L.}~\bibnamefont
  {Jaime}}, \bibinfo {author} {\bibfnamefont {M.}~\bibnamefont {Salgado}}, \
  and\ \bibinfo {author} {\bibfnamefont {L.}~\bibnamefont {Patino}},\ }\href
  {\doibase 10.1007/978-3-319-06761-2\_51} {\bibfield  {journal} {\bibinfo
  {journal} {Springer Proc. Phys.}\ }\textbf {\bibinfo {volume} {157}},\
  \bibinfo {pages} {363} (\bibinfo {year} {2014})},\ \Eprint
  {http://arxiv.org/abs/1211.0015} {arXiv:1211.0015 [gr-qc]} \BibitemShut
  {NoStop}%
\bibitem [{\citenamefont {Maartens}\ and\ \citenamefont
  {Koyama}(2010)}]{Maartens:2010ar}%
  \BibitemOpen
  \bibfield  {author} {\bibinfo {author} {\bibfnamefont {R.}~\bibnamefont
  {Maartens}}\ and\ \bibinfo {author} {\bibfnamefont {K.}~\bibnamefont
  {Koyama}},\ }\href {\doibase 10.12942/lrr-2010-5} {\bibfield  {journal}
  {\bibinfo  {journal} {Living Rev. Rel.}\ }\textbf {\bibinfo {volume} {13}},\
  \bibinfo {pages} {5} (\bibinfo {year} {2010})},\ \Eprint
  {http://arxiv.org/abs/1004.3962} {arXiv:1004.3962 [hep-th]} \BibitemShut
  {NoStop}%
\bibitem [{\citenamefont {Garcia-Aspeitia}\ \emph {et~al.}(2018)\citenamefont
  {Garcia-Aspeitia}, \citenamefont {Hernandez-Almada}, \citenamefont
  {Maga\~na}, \citenamefont {Amante}, \citenamefont {Motta},\ and\
  \citenamefont {Mart\'\i{}nez-Robles}}]{Garcia-Aspeitia:2018fvw}%
  \BibitemOpen
  \bibfield  {author} {\bibinfo {author} {\bibfnamefont {M.~A.}\ \bibnamefont
  {Garcia-Aspeitia}}, \bibinfo {author} {\bibfnamefont {A.}~\bibnamefont
  {Hernandez-Almada}}, \bibinfo {author} {\bibfnamefont {J.}~\bibnamefont
  {Maga\~na}}, \bibinfo {author} {\bibfnamefont {M.~H.}\ \bibnamefont
  {Amante}}, \bibinfo {author} {\bibfnamefont {V.}~\bibnamefont {Motta}}, \
  and\ \bibinfo {author} {\bibfnamefont {C.}~\bibnamefont
  {Mart\'\i{}nez-Robles}},\ }\href {\doibase 10.1103/PhysRevD.97.101301}
  {\bibfield  {journal} {\bibinfo  {journal} {Phys. Rev. D}\ }\textbf {\bibinfo
  {volume} {97}},\ \bibinfo {pages} {101301} (\bibinfo {year} {2018})},\
  \Eprint {http://arxiv.org/abs/1804.05085} {arXiv:1804.05085 [gr-qc]}
  \BibitemShut {NoStop}%
\bibitem [{\citenamefont {Josset}\ \emph {et~al.}(2017)\citenamefont {Josset},
  \citenamefont {Perez},\ and\ \citenamefont {Sudarsky}}]{Josset:2016}%
  \BibitemOpen
  \bibfield  {author} {\bibinfo {author} {\bibfnamefont {T.}~\bibnamefont
  {Josset}}, \bibinfo {author} {\bibfnamefont {A.}~\bibnamefont {Perez}}, \
  and\ \bibinfo {author} {\bibfnamefont {D.}~\bibnamefont {Sudarsky}},\ }\href
  {\doibase 10.1103/PhysRevLett.118.021102} {\bibfield  {journal} {\bibinfo
  {journal} {Phys. Rev. Lett.}\ }\textbf {\bibinfo {volume} {118}},\ \bibinfo
  {pages} {021102} (\bibinfo {year} {2017})},\ \Eprint
  {http://arxiv.org/abs/1604.04183} {arXiv:1604.04183 [gr-qc]} \BibitemShut
  {NoStop}%
\bibitem [{\citenamefont {Garc\'ia-Aspeitia}\ \emph {et~al.}(2019)\citenamefont
  {Garc\'ia-Aspeitia}, \citenamefont {Mart\'inez-Robles}, \citenamefont
  {Hern\'andez-Almada}, \citenamefont {Maga\~na},\ and\ \citenamefont
  {Motta}}]{Garcia-Aspeitia:2019yni}%
  \BibitemOpen
  \bibfield  {author} {\bibinfo {author} {\bibfnamefont {M.~A.}\ \bibnamefont
  {Garc\'ia-Aspeitia}}, \bibinfo {author} {\bibfnamefont {C.}~\bibnamefont
  {Mart\'inez-Robles}}, \bibinfo {author} {\bibfnamefont {A.}~\bibnamefont
  {Hern\'andez-Almada}}, \bibinfo {author} {\bibfnamefont {J.}~\bibnamefont
  {Maga\~na}}, \ and\ \bibinfo {author} {\bibfnamefont {V.}~\bibnamefont
  {Motta}},\ }\href {\doibase 10.1103/PhysRevD.99.123525} {\bibfield  {journal}
  {\bibinfo  {journal} {Phys. Rev.}\ }\textbf {\bibinfo {volume} {D99}},\
  \bibinfo {pages} {123525} (\bibinfo {year} {2019})},\ \Eprint
  {http://arxiv.org/abs/1903.06344} {arXiv:1903.06344 [gr-qc]} \BibitemShut
  {NoStop}%
\bibitem [{\citenamefont {Garc\'\i{}a-Aspeitia}\ \emph
  {et~al.}(2019)\citenamefont {Garc\'\i{}a-Aspeitia}, \citenamefont
  {Hern\'andez-Almada}, \citenamefont {Maga\~na},\ and\ \citenamefont
  {Motta}}]{Garcia-Aspeitia:2019yod}%
  \BibitemOpen
  \bibfield  {author} {\bibinfo {author} {\bibfnamefont {M.~A.}\ \bibnamefont
  {Garc\'\i{}a-Aspeitia}}, \bibinfo {author} {\bibfnamefont {A.}~\bibnamefont
  {Hern\'andez-Almada}}, \bibinfo {author} {\bibfnamefont {J.}~\bibnamefont
  {Maga\~na}}, \ and\ \bibinfo {author} {\bibfnamefont {V.}~\bibnamefont
  {Motta}},\ }\href@noop {} {\  (\bibinfo {year} {2019})},\ \Eprint
  {http://arxiv.org/abs/1912.07500} {arXiv:1912.07500 [astro-ph.CO]}
  \BibitemShut {NoStop}%
\bibitem [{\citenamefont {Corral}\ \emph {et~al.}(2020)\citenamefont {Corral},
  \citenamefont {Cruz},\ and\ \citenamefont {Gonz\'alez}}]{Corral:2020lxt}%
  \BibitemOpen
  \bibfield  {author} {\bibinfo {author} {\bibfnamefont {C.}~\bibnamefont
  {Corral}}, \bibinfo {author} {\bibfnamefont {N.}~\bibnamefont {Cruz}}, \ and\
  \bibinfo {author} {\bibfnamefont {E.}~\bibnamefont {Gonz\'alez}},\ }\href
  {\doibase 10.1103/PhysRevD.102.023508} {\bibfield  {journal} {\bibinfo
  {journal} {Phys. Rev. D}\ }\textbf {\bibinfo {volume} {102}},\ \bibinfo
  {pages} {023508} (\bibinfo {year} {2020})},\ \Eprint
  {http://arxiv.org/abs/2005.06052} {arXiv:2005.06052 [gr-qc]} \BibitemShut
  {NoStop}%
\bibitem [{\citenamefont {Li}\ and\ \citenamefont {Shafieloo}(2019)}]{Li_2019}%
  \BibitemOpen
  \bibfield  {author} {\bibinfo {author} {\bibfnamefont {X.}~\bibnamefont
  {Li}}\ and\ \bibinfo {author} {\bibfnamefont {A.}~\bibnamefont {Shafieloo}},\
  }\href {\doibase 10.3847/2041-8213/ab3e09} {\bibfield  {journal} {\bibinfo
  {journal} {The Astrophysical Journal}\ }\textbf {\bibinfo {volume} {883}},\
  \bibinfo {pages} {L3} (\bibinfo {year} {2019})}\BibitemShut {NoStop}%
\bibitem [{\citenamefont {Li}\ and\ \citenamefont
  {Shafieloo}(2020)}]{Li:2020ybr}%
  \BibitemOpen
  \bibfield  {author} {\bibinfo {author} {\bibfnamefont {X.}~\bibnamefont
  {Li}}\ and\ \bibinfo {author} {\bibfnamefont {A.}~\bibnamefont {Shafieloo}},\
  }\href {\doibase 10.3847/1538-4357/abb3d0} {\bibfield  {journal} {\bibinfo
  {journal} {Astrophys. J.}\ }\textbf {\bibinfo {volume} {902}},\ \bibinfo
  {pages} {58} (\bibinfo {year} {2020})},\ \Eprint
  {http://arxiv.org/abs/2001.05103} {arXiv:2001.05103 [astro-ph.CO]}
  \BibitemShut {NoStop}%
\bibitem [{\citenamefont {Hern\'andez-Almada}\ \emph
  {et~al.}(2020{\natexlab{a}})\citenamefont {Hern\'andez-Almada}, \citenamefont
  {Leon}, \citenamefont {Maga\~na}, \citenamefont {Garc\'\i{}a-Aspeitia},\ and\
  \citenamefont {Motta}}]{Hernandez-Almada:2020uyr}%
  \BibitemOpen
  \bibfield  {author} {\bibinfo {author} {\bibfnamefont {A.}~\bibnamefont
  {Hern\'andez-Almada}}, \bibinfo {author} {\bibfnamefont {G.}~\bibnamefont
  {Leon}}, \bibinfo {author} {\bibfnamefont {J.}~\bibnamefont {Maga\~na}},
  \bibinfo {author} {\bibfnamefont {M.~A.}\ \bibnamefont
  {Garc\'\i{}a-Aspeitia}}, \ and\ \bibinfo {author} {\bibfnamefont
  {V.}~\bibnamefont {Motta}},\ }\href {\doibase 10.1093/mnras/staa2052}
  {\bibfield  {journal} {\bibinfo  {journal} {Mon. Not. Roy. Astron. Soc.}\
  }\textbf {\bibinfo {volume} {497}},\ \bibinfo {pages} {1590} (\bibinfo {year}
  {2020}{\natexlab{a}})},\ \Eprint {http://arxiv.org/abs/2002.12881}
  {arXiv:2002.12881 [astro-ph.CO]} \BibitemShut {NoStop}%
\bibitem [{\citenamefont {Copeland}\ \emph {et~al.}(2006)\citenamefont
  {Copeland}, \citenamefont {Sami},\ and\ \citenamefont
  {Tsujikawa}}]{Copeland:2006wr}%
  \BibitemOpen
  \bibfield  {author} {\bibinfo {author} {\bibfnamefont {E.~J.}\ \bibnamefont
  {Copeland}}, \bibinfo {author} {\bibfnamefont {M.}~\bibnamefont {Sami}}, \
  and\ \bibinfo {author} {\bibfnamefont {S.}~\bibnamefont {Tsujikawa}},\ }\href
  {\doibase 10.1142/S021827180600942X} {\bibfield  {journal} {\bibinfo
  {journal} {Int. J. Mod. Phys.}\ }\textbf {\bibinfo {volume} {D15}},\ \bibinfo
  {pages} {1753} (\bibinfo {year} {2006})},\ \Eprint
  {http://arxiv.org/abs/hep-th/0603057} {arXiv:hep-th/0603057 [hep-th]}
  \BibitemShut {NoStop}%
\bibitem [{\citenamefont {Cruz}\ \emph
  {et~al.}(2019{\natexlab{a}})\citenamefont {Cruz}, \citenamefont
  {Hern\'andez-Almada},\ and\ \citenamefont {Cornejo-P\'erez}}]{Cruz:2019wbl}%
  \BibitemOpen
  \bibfield  {author} {\bibinfo {author} {\bibfnamefont {N.}~\bibnamefont
  {Cruz}}, \bibinfo {author} {\bibfnamefont {A.}~\bibnamefont
  {Hern\'andez-Almada}}, \ and\ \bibinfo {author} {\bibfnamefont
  {O.}~\bibnamefont {Cornejo-P\'erez}},\ }\href {\doibase
  10.1103/PhysRevD.100.083524} {\bibfield  {journal} {\bibinfo  {journal}
  {Phys. Rev. D}\ }\textbf {\bibinfo {volume} {100}},\ \bibinfo {pages}
  {083524} (\bibinfo {year} {2019}{\natexlab{a}})},\ \Eprint
  {http://arxiv.org/abs/1909.12418} {arXiv:1909.12418 [astro-ph.CO]}
  \BibitemShut {NoStop}%
\bibitem [{\citenamefont {Cruz}\ \emph
  {et~al.}(2017{\natexlab{a}})\citenamefont {Cruz}, \citenamefont {Cruz},\ and\
  \citenamefont {Lepe}}]{Cruz:2017bcv}%
  \BibitemOpen
  \bibfield  {author} {\bibinfo {author} {\bibfnamefont {M.}~\bibnamefont
  {Cruz}}, \bibinfo {author} {\bibfnamefont {N.}~\bibnamefont {Cruz}}, \ and\
  \bibinfo {author} {\bibfnamefont {S.}~\bibnamefont {Lepe}},\ }\href {\doibase
  10.1103/PhysRevD.96.124020} {\bibfield  {journal} {\bibinfo  {journal} {Phys.
  Rev. D}\ }\textbf {\bibinfo {volume} {96}},\ \bibinfo {pages} {124020}
  (\bibinfo {year} {2017}{\natexlab{a}})},\ \Eprint
  {http://arxiv.org/abs/1710.02607} {arXiv:1710.02607 [gr-qc]} \BibitemShut
  {NoStop}%
\bibitem [{\citenamefont {Brevik}\ and\ \citenamefont
  {Gorbunova}(2005{\natexlab{a}})}]{Brevik:2005bj}%
  \BibitemOpen
  \bibfield  {author} {\bibinfo {author} {\bibfnamefont {I.~H.}\ \bibnamefont
  {Brevik}}\ and\ \bibinfo {author} {\bibfnamefont {O.}~\bibnamefont
  {Gorbunova}},\ }\href {\doibase 10.1007/s10714-005-0178-9} {\bibfield
  {journal} {\bibinfo  {journal} {Gen. Rel. Grav.}\ }\textbf {\bibinfo {volume}
  {37}},\ \bibinfo {pages} {2039} (\bibinfo {year} {2005}{\natexlab{a}})},\
  \Eprint {http://arxiv.org/abs/gr-qc/0504001} {arXiv:gr-qc/0504001}
  \BibitemShut {NoStop}%
\bibitem [{\citenamefont {Chaplygin}(1904)}]{Chaplygin}%
  \BibitemOpen
  \bibfield  {author} {\bibinfo {author} {\bibfnamefont {S.~A.}\ \bibnamefont
  {Chaplygin}},\ }\href@noop {} {\bibfield  {journal} {\bibinfo  {journal}
  {Sci. Mem. Moscow Univ. Math. Phys.}\ }\textbf {\bibinfo {volume} {21}}
  (\bibinfo {year} {1904})}\BibitemShut {NoStop}%
\bibitem [{\citenamefont {Kamenshchik}\ \emph
  {et~al.}(2001{\natexlab{a}})\citenamefont {Kamenshchik}, \citenamefont
  {Moschella},\ and\ \citenamefont {Pasquier}}]{Kamenshchik:2001cp}%
  \BibitemOpen
  \bibfield  {author} {\bibinfo {author} {\bibfnamefont {A.~Y.}\ \bibnamefont
  {Kamenshchik}}, \bibinfo {author} {\bibfnamefont {U.}~\bibnamefont
  {Moschella}}, \ and\ \bibinfo {author} {\bibfnamefont {V.}~\bibnamefont
  {Pasquier}},\ }\href {\doibase 10.1016/S0370-2693(01)00571-8} {\bibfield
  {journal} {\bibinfo  {journal} {Phys. Lett. B}\ }\textbf {\bibinfo {volume}
  {511}},\ \bibinfo {pages} {265} (\bibinfo {year} {2001}{\natexlab{a}})},\
  \Eprint {http://arxiv.org/abs/gr-qc/0103004} {arXiv:gr-qc/0103004}
  \BibitemShut {NoStop}%
\bibitem [{\citenamefont {Kamenshchik}\ \emph
  {et~al.}(2001{\natexlab{b}})\citenamefont {Kamenshchik}, \citenamefont
  {Moschella},\ and\ \citenamefont {Pasquier}}]{Kamenshchik}%
  \BibitemOpen
  \bibfield  {author} {\bibinfo {author} {\bibfnamefont {A.}~\bibnamefont
  {Kamenshchik}}, \bibinfo {author} {\bibfnamefont {U.}~\bibnamefont
  {Moschella}}, \ and\ \bibinfo {author} {\bibfnamefont {V.}~\bibnamefont
  {Pasquier}},\ }\href {\doibase https://doi.org/10.1016/S0370-2693(01)00571-8}
  {\bibfield  {journal} {\bibinfo  {journal} {Physics Letters B}\ }\textbf
  {\bibinfo {volume} {511}},\ \bibinfo {pages} {265 } (\bibinfo {year}
  {2001}{\natexlab{b}})}\BibitemShut {NoStop}%
\bibitem [{\citenamefont {Bilić}\ \emph {et~al.}(2002)\citenamefont {Bilić},
  \citenamefont {Tupper},\ and\ \citenamefont {Viollier}}]{Bilic}%
  \BibitemOpen
  \bibfield  {author} {\bibinfo {author} {\bibfnamefont {N.}~\bibnamefont
  {Bilić}}, \bibinfo {author} {\bibfnamefont {G.~B.}\ \bibnamefont {Tupper}},
  \ and\ \bibinfo {author} {\bibfnamefont {R.~D.}\ \bibnamefont {Viollier}},\
  }\href {\doibase https://doi.org/10.1016/S0370-2693(02)01716-1} {\bibfield
  {journal} {\bibinfo  {journal} {Physics Letters B}\ }\textbf {\bibinfo
  {volume} {535}},\ \bibinfo {pages} {17 } (\bibinfo {year}
  {2002})}\BibitemShut {NoStop}%
\bibitem [{\citenamefont {Fabris}\ \emph {et~al.}(2002)\citenamefont {Fabris},
  \citenamefont {Gonçalves},\ and\ \citenamefont {de~Souza}}]{Fabris}%
  \BibitemOpen
  \bibfield  {author} {\bibinfo {author} {\bibfnamefont {J.~C.}\ \bibnamefont
  {Fabris}}, \bibinfo {author} {\bibfnamefont {S.~V.~B.}\ \bibnamefont
  {Gonçalves}}, \ and\ \bibinfo {author} {\bibfnamefont {P.~E.}\ \bibnamefont
  {de~Souza}},\ }\href {\doibase 10.1023/A:1015266421750} {\bibfield  {journal}
  {\bibinfo  {journal} {General Relativity and Gravitation}\ }\textbf {\bibinfo
  {volume} {34}},\ \bibinfo {pages} {53} (\bibinfo {year} {2002})}\BibitemShut
  {NoStop}%
\bibitem [{\citenamefont {Hernandez-Almada}\ \emph {et~al.}(2019)\citenamefont
  {Hernandez-Almada}, \citenamefont {Magana}, \citenamefont {Garcia-Aspeitia},\
  and\ \citenamefont {Motta}}]{Hernandez-Almada:2018osh}%
  \BibitemOpen
  \bibfield  {author} {\bibinfo {author} {\bibfnamefont {A.}~\bibnamefont
  {Hernandez-Almada}}, \bibinfo {author} {\bibfnamefont {J.}~\bibnamefont
  {Magana}}, \bibinfo {author} {\bibfnamefont {M.~A.}\ \bibnamefont
  {Garcia-Aspeitia}}, \ and\ \bibinfo {author} {\bibfnamefont {V.}~\bibnamefont
  {Motta}},\ }\href {\doibase 10.1140/epjc/s10052-018-6521-6} {\bibfield
  {journal} {\bibinfo  {journal} {Eur. Phys. J. C}\ }\textbf {\bibinfo {volume}
  {79}},\ \bibinfo {pages} {12} (\bibinfo {year} {2019})},\ \Eprint
  {http://arxiv.org/abs/1805.07895} {arXiv:1805.07895 [astro-ph.CO]}
  \BibitemShut {NoStop}%
\bibitem [{\citenamefont {Villanueva}(2015)}]{Villanueva_2015}%
  \BibitemOpen
  \bibfield  {author} {\bibinfo {author} {\bibfnamefont {J.}~\bibnamefont
  {Villanueva}},\ }\href {\doibase 10.1088/1475-7516/2015/07/045} {\bibfield
  {journal} {\bibinfo  {journal} {Journal of Cosmology and Astroparticle
  Physics}\ }\textbf {\bibinfo {volume} {2015}},\ \bibinfo {pages} {045}
  (\bibinfo {year} {2015})}\BibitemShut {NoStop}%
\bibitem [{\citenamefont {Sandvik}\ \emph {et~al.}(2004)\citenamefont
  {Sandvik}, \citenamefont {Tegmark}, \citenamefont {Zaldarriaga},\ and\
  \citenamefont {Waga}}]{Sandvik:2004}%
  \BibitemOpen
  \bibfield  {author} {\bibinfo {author} {\bibfnamefont {H.~B.}\ \bibnamefont
  {Sandvik}}, \bibinfo {author} {\bibfnamefont {M.}~\bibnamefont {Tegmark}},
  \bibinfo {author} {\bibfnamefont {M.}~\bibnamefont {Zaldarriaga}}, \ and\
  \bibinfo {author} {\bibfnamefont {I.}~\bibnamefont {Waga}},\ }\href {\doibase
  10.1103/PhysRevD.69.123524} {\bibfield  {journal} {\bibinfo  {journal} {Phys.
  Rev. D}\ }\textbf {\bibinfo {volume} {69}},\ \bibinfo {pages} {123524}
  (\bibinfo {year} {2004})}\BibitemShut {NoStop}%
\bibitem [{\citenamefont {Perrotta}\ \emph {et~al.}(2004)\citenamefont
  {Perrotta}, \citenamefont {Matarrese},\ and\ \citenamefont
  {Torki}}]{Perrotta:2004}%
  \BibitemOpen
  \bibfield  {author} {\bibinfo {author} {\bibfnamefont {F.}~\bibnamefont
  {Perrotta}}, \bibinfo {author} {\bibfnamefont {S.}~\bibnamefont {Matarrese}},
  \ and\ \bibinfo {author} {\bibfnamefont {M.}~\bibnamefont {Torki}},\ }\href
  {\doibase 10.1103/PhysRevD.70.121304} {\bibfield  {journal} {\bibinfo
  {journal} {Phys. Rev. D}\ }\textbf {\bibinfo {volume} {70}},\ \bibinfo
  {pages} {121304} (\bibinfo {year} {2004})}\BibitemShut {NoStop}%
\bibitem [{\citenamefont {Eckart}(1940)}]{Eckart}%
  \BibitemOpen
  \bibfield  {author} {\bibinfo {author} {\bibfnamefont {C.}~\bibnamefont
  {Eckart}},\ }\href {\doibase 10.1103/PhysRev.58.919} {\bibfield  {journal}
  {\bibinfo  {journal} {Phys. Rev.}\ }\textbf {\bibinfo {volume} {58}},\
  \bibinfo {pages} {919} (\bibinfo {year} {1940})}\BibitemShut {NoStop}%
\bibitem [{\citenamefont {Israel}\ and\ \citenamefont
  {Stewart}(1979)}]{Israel}%
  \BibitemOpen
  \bibfield  {author} {\bibinfo {author} {\bibfnamefont {W.}~\bibnamefont
  {Israel}}\ and\ \bibinfo {author} {\bibfnamefont {J.}~\bibnamefont
  {Stewart}},\ }\href {\doibase https://doi.org/10.1016/0003-4916(79)90130-1}
  {\bibfield  {journal} {\bibinfo  {journal} {Annals of Physics}\ }\textbf
  {\bibinfo {volume} {118}},\ \bibinfo {pages} {341 } (\bibinfo {year}
  {1979})}\BibitemShut {NoStop}%
\bibitem [{\citenamefont {Cruz}\ \emph
  {et~al.}(2017{\natexlab{b}})\citenamefont {Cruz}, \citenamefont {Cruz},\ and\
  \citenamefont {Lepe}}]{Cruz}%
  \BibitemOpen
  \bibfield  {author} {\bibinfo {author} {\bibfnamefont {M.}~\bibnamefont
  {Cruz}}, \bibinfo {author} {\bibfnamefont {N.}~\bibnamefont {Cruz}}, \ and\
  \bibinfo {author} {\bibfnamefont {S.}~\bibnamefont {Lepe}},\ }\href {\doibase
  10.1103/PhysRevD.96.124020} {\bibfield  {journal} {\bibinfo  {journal} {Phys.
  Rev. D}\ }\textbf {\bibinfo {volume} {96}},\ \bibinfo {pages} {124020}
  (\bibinfo {year} {2017}{\natexlab{b}})}\BibitemShut {NoStop}%
\bibitem [{\citenamefont {Cruz}\ \emph
  {et~al.}(2017{\natexlab{c}})\citenamefont {Cruz}, \citenamefont {Cruz},\ and\
  \citenamefont {Lepe}}]{CRUZ2017159}%
  \BibitemOpen
  \bibfield  {author} {\bibinfo {author} {\bibfnamefont {M.}~\bibnamefont
  {Cruz}}, \bibinfo {author} {\bibfnamefont {N.}~\bibnamefont {Cruz}}, \ and\
  \bibinfo {author} {\bibfnamefont {S.}~\bibnamefont {Lepe}},\ }\href {\doibase
  https://doi.org/10.1016/j.physletb.2017.03.065} {\bibfield  {journal}
  {\bibinfo  {journal} {Physics Letters B}\ }\textbf {\bibinfo {volume}
  {769}},\ \bibinfo {pages} {159 } (\bibinfo {year}
  {2017}{\natexlab{c}})}\BibitemShut {NoStop}%
\bibitem [{\citenamefont {Cruz}\ \emph {et~al.}(2020)\citenamefont {Cruz},
  \citenamefont {Gonz\'alez},\ and\ \citenamefont {Palma}}]{Cruz:2018psw}%
  \BibitemOpen
  \bibfield  {author} {\bibinfo {author} {\bibfnamefont {N.}~\bibnamefont
  {Cruz}}, \bibinfo {author} {\bibfnamefont {E.}~\bibnamefont {Gonz\'alez}}, \
  and\ \bibinfo {author} {\bibfnamefont {G.}~\bibnamefont {Palma}},\ }\href
  {\doibase 10.1007/s10714-020-02712-z} {\bibfield  {journal} {\bibinfo
  {journal} {Gen. Rel. Grav.}\ }\textbf {\bibinfo {volume} {52}},\ \bibinfo
  {pages} {62} (\bibinfo {year} {2020})},\ \Eprint
  {http://arxiv.org/abs/1812.05009} {arXiv:1812.05009 [gr-qc]} \BibitemShut
  {NoStop}%
\bibitem [{\citenamefont {Cruz}\ \emph
  {et~al.}(2019{\natexlab{b}})\citenamefont {Cruz}, \citenamefont
  {Hern\'andez-Almada},\ and\ \citenamefont
  {Cornejo-P\'erez}}]{CruzyHernandez}%
  \BibitemOpen
  \bibfield  {author} {\bibinfo {author} {\bibfnamefont {N.}~\bibnamefont
  {Cruz}}, \bibinfo {author} {\bibfnamefont {A.}~\bibnamefont
  {Hern\'andez-Almada}}, \ and\ \bibinfo {author} {\bibfnamefont
  {O.}~\bibnamefont {Cornejo-P\'erez}},\ }\href {\doibase
  10.1103/PhysRevD.100.083524} {\bibfield  {journal} {\bibinfo  {journal}
  {Phys. Rev. D}\ }\textbf {\bibinfo {volume} {100}},\ \bibinfo {pages}
  {083524} (\bibinfo {year} {2019}{\natexlab{b}})}\BibitemShut {NoStop}%
\bibitem [{\citenamefont {Cruz}\ \emph
  {et~al.}(2019{\natexlab{c}})\citenamefont {Cruz}, \citenamefont {Cruz},\ and\
  \citenamefont {Lepe}}]{Lepe}%
  \BibitemOpen
  \bibfield  {author} {\bibinfo {author} {\bibfnamefont {M.}~\bibnamefont
  {Cruz}}, \bibinfo {author} {\bibfnamefont {N.}~\bibnamefont {Cruz}}, \ and\
  \bibinfo {author} {\bibfnamefont {S.}~\bibnamefont {Lepe}},\ }\href@noop {}
  {\  (\bibinfo {year} {2019}{\natexlab{c}})},\ \Eprint
  {http://arxiv.org/abs/1911.04539} {arXiv:1911.04539 [gr-qc]} \BibitemShut
  {NoStop}%
\bibitem [{\citenamefont {Murphy}(1973)}]{Murphy}%
  \BibitemOpen
  \bibfield  {author} {\bibinfo {author} {\bibfnamefont {G.~L.}\ \bibnamefont
  {Murphy}},\ }\href {\doibase 10.1103/PhysRevD.8.4231} {\bibfield  {journal}
  {\bibinfo  {journal} {Phys. Rev. D}\ }\textbf {\bibinfo {volume} {8}},\
  \bibinfo {pages} {4231} (\bibinfo {year} {1973})}\BibitemShut {NoStop}%
\bibitem [{\citenamefont {Padmanabhan}\ and\ \citenamefont
  {Chitre}(1987)}]{Padma}%
  \BibitemOpen
  \bibfield  {author} {\bibinfo {author} {\bibfnamefont {T.}~\bibnamefont
  {Padmanabhan}}\ and\ \bibinfo {author} {\bibfnamefont {S.}~\bibnamefont
  {Chitre}},\ }\href {\doibase https://doi.org/10.1016/0375-9601(87)90104-6}
  {\bibfield  {journal} {\bibinfo  {journal} {Physics Letters A}\ }\textbf
  {\bibinfo {volume} {120}},\ \bibinfo {pages} {433 } (\bibinfo {year}
  {1987})}\BibitemShut {NoStop}%
\bibitem [{\citenamefont {Brevik}\ and\ \citenamefont
  {Gorbunova}(2005{\natexlab{b}})}]{Brevik}%
  \BibitemOpen
  \bibfield  {author} {\bibinfo {author} {\bibfnamefont {I.}~\bibnamefont
  {Brevik}}\ and\ \bibinfo {author} {\bibfnamefont {O.}~\bibnamefont
  {Gorbunova}},\ }\href {\doibase 10.1007/s10714-005-0178-9} {\bibfield
  {journal} {\bibinfo  {journal} {General Relativity and Gravitation}\ }\textbf
  {\bibinfo {volume} {37}},\ \bibinfo {pages} {2039} (\bibinfo {year}
  {2005}{\natexlab{b}})}\BibitemShut {NoStop}%
\bibitem [{\citenamefont {Xin-He}\ and\ \citenamefont {Xu}(2009)}]{Xin_He}%
  \BibitemOpen
  \bibfield  {author} {\bibinfo {author} {\bibfnamefont {M.}~\bibnamefont
  {Xin-He}}\ and\ \bibinfo {author} {\bibfnamefont {D.}~\bibnamefont {Xu}},\
  }\href {\doibase 10.1088/0253-6102/52/2/36} {\bibfield  {journal} {\bibinfo
  {journal} {Communications in Theoretical Physics}\ }\textbf {\bibinfo
  {volume} {52}},\ \bibinfo {pages} {377} (\bibinfo {year} {2009})}\BibitemShut
  {NoStop}%
\bibitem [{\citenamefont {Avelino}\ and\ \citenamefont
  {Nucamendi}(2010)}]{Avelino}%
  \BibitemOpen
  \bibfield  {author} {\bibinfo {author} {\bibfnamefont {A.}~\bibnamefont
  {Avelino}}\ and\ \bibinfo {author} {\bibfnamefont {U.}~\bibnamefont
  {Nucamendi}},\ }\href {\doibase 10.1088/1475-7516/2010/08/009} {\bibfield
  {journal} {\bibinfo  {journal} {Journal of Cosmology and Astroparticle
  Physics}\ }\textbf {\bibinfo {volume} {2010}},\ \bibinfo {pages} {009}
  (\bibinfo {year} {2010})}\BibitemShut {NoStop}%
\bibitem [{\citenamefont {Hernández-Almada}(2019)}]{AlmadaViscoso}%
  \BibitemOpen
  \bibfield  {author} {\bibinfo {author} {\bibfnamefont {A.}~\bibnamefont
  {Hernández-Almada}},\ }\href {\doibase 10.1140/epjc/s10052-019-7264-8}
  {\bibfield  {journal} {\bibinfo  {journal} {The European Physical Journal C}\
  }\textbf {\bibinfo {volume} {79}},\ \bibinfo {pages} {751} (\bibinfo {year}
  {2019})}\BibitemShut {NoStop}%
\bibitem [{\citenamefont {Folomeev}\ and\ \citenamefont
  {Gurovich}(2008)}]{Folomeev}%
  \BibitemOpen
  \bibfield  {author} {\bibinfo {author} {\bibfnamefont {V.}~\bibnamefont
  {Folomeev}}\ and\ \bibinfo {author} {\bibfnamefont {V.}~\bibnamefont
  {Gurovich}},\ }\href {\doibase
  https://doi.org/10.1016/j.physletb.2008.01.068} {\bibfield  {journal}
  {\bibinfo  {journal} {Physics Letters B}\ }\textbf {\bibinfo {volume}
  {661}},\ \bibinfo {pages} {75 } (\bibinfo {year} {2008})}\BibitemShut
  {NoStop}%
\bibitem [{\citenamefont {Hern\'andez-Almada}\ \emph
  {et~al.}(2020{\natexlab{b}})\citenamefont {Hern\'andez-Almada}, \citenamefont
  {Garc\'{\i}a-Aspeitia}, \citenamefont {Maga\~na},\ and\ \citenamefont
  {Motta}}]{Almada:2020}%
  \BibitemOpen
  \bibfield  {author} {\bibinfo {author} {\bibfnamefont {A.}~\bibnamefont
  {Hern\'andez-Almada}}, \bibinfo {author} {\bibfnamefont {M.~A.}\ \bibnamefont
  {Garc\'{\i}a-Aspeitia}}, \bibinfo {author} {\bibfnamefont {J.}~\bibnamefont
  {Maga\~na}}, \ and\ \bibinfo {author} {\bibfnamefont {V.}~\bibnamefont
  {Motta}},\ }\href {\doibase 10.1103/PhysRevD.101.063516} {\bibfield
  {journal} {\bibinfo  {journal} {Phys. Rev. D}\ }\textbf {\bibinfo {volume}
  {101}},\ \bibinfo {pages} {063516} (\bibinfo {year}
  {2020}{\natexlab{b}})}\BibitemShut {NoStop}%
\bibitem [{\citenamefont {Zhai}\ \emph {et~al.}(2006)\citenamefont {Zhai},
  \citenamefont {Xu},\ and\ \citenamefont {Li}}]{ZHAI:2006}%
  \BibitemOpen
  \bibfield  {author} {\bibinfo {author} {\bibfnamefont {X.-H.}\ \bibnamefont
  {Zhai}}, \bibinfo {author} {\bibfnamefont {Y.-D.}\ \bibnamefont {Xu}}, \ and\
  \bibinfo {author} {\bibfnamefont {X.-Z.}\ \bibnamefont {Li}},\ }\href
  {\doibase 10.1142/S0218271806008784} {\bibfield  {journal} {\bibinfo
  {journal} {Int. J. Mod. Phys. D}\ }\textbf {\bibinfo {volume} {15}},\
  \bibinfo {pages} {1151} (\bibinfo {year} {2006})},\ \Eprint
  {http://arxiv.org/abs/astro-ph/0511814} {arXiv:astro-ph/0511814} \BibitemShut
  {NoStop}%
\bibitem [{\citenamefont {Xu}\ \emph {et~al.}(2012)\citenamefont {Xu},
  \citenamefont {Huang},\ and\ \citenamefont {Zhai}}]{Xu:2012}%
  \BibitemOpen
  \bibfield  {author} {\bibinfo {author} {\bibfnamefont {Y.}~\bibnamefont
  {Xu}}, \bibinfo {author} {\bibfnamefont {Z.}~\bibnamefont {Huang}}, \ and\
  \bibinfo {author} {\bibfnamefont {X.}~\bibnamefont {Zhai}},\ }\href {\doibase
  10.1007/s10509-011-0850-3} {\bibfield  {journal} {\bibinfo  {journal}
  {Astrophys Space Sci}\ }\textbf {\bibinfo {volume} {337}},\ \bibinfo {pages}
  {493} (\bibinfo {year} {2012})}\BibitemShut {NoStop}%
\bibitem [{\citenamefont {Saadat}\ and\ \citenamefont
  {Pourhassan}(2013{\natexlab{a}})}]{Saadat2:2013}%
  \BibitemOpen
  \bibfield  {author} {\bibinfo {author} {\bibfnamefont {H.}~\bibnamefont
  {Saadat}}\ and\ \bibinfo {author} {\bibfnamefont {B.}~\bibnamefont
  {Pourhassan}},\ }\href {\doibase 10.1007/s10509-012-1301-5} {\bibfield
  {journal} {\bibinfo  {journal} {Astrophys Space Sci}\ }\textbf {\bibinfo
  {volume} {344}},\ \bibinfo {pages} {237} (\bibinfo {year}
  {2013}{\natexlab{a}})}\BibitemShut {NoStop}%
\bibitem [{\citenamefont {Li}\ and\ \citenamefont {Xu}(2013)}]{Li-Xu:2013}%
  \BibitemOpen
  \bibfield  {author} {\bibinfo {author} {\bibfnamefont {W.}~\bibnamefont
  {Li}}\ and\ \bibinfo {author} {\bibfnamefont {L.}~\bibnamefont {Xu}},\ }\href
  {\doibase 10.1140/epjc/s10052-013-2471-1} {\bibfield  {journal} {\bibinfo
  {journal} {Eur. Phys. J. C}\ }\textbf {\bibinfo {volume} {73}},\ \bibinfo
  {pages} {2471} (\bibinfo {year} {2013})}\BibitemShut {NoStop}%
\bibitem [{\citenamefont {Saadat}\ and\ \citenamefont
  {Pourhassan}(2013{\natexlab{b}})}]{Saadat:2013}%
  \BibitemOpen
  \bibfield  {author} {\bibinfo {author} {\bibfnamefont {H.}~\bibnamefont
  {Saadat}}\ and\ \bibinfo {author} {\bibfnamefont {B.}~\bibnamefont
  {Pourhassan}},\ }\href {\doibase 10.1007/s10509-012-1268-2} {\bibfield
  {journal} {\bibinfo  {journal} {Astrophys Space Sci}\ }\textbf {\bibinfo
  {volume} {343}},\ \bibinfo {pages} {783–786} (\bibinfo {year}
  {2013}{\natexlab{b}})}\BibitemShut {NoStop}%
\bibitem [{\citenamefont {Jawad}\ and\ \citenamefont
  {Iqbal}(2016)}]{Jawad:2016dcp}%
  \BibitemOpen
  \bibfield  {author} {\bibinfo {author} {\bibfnamefont {A.}~\bibnamefont
  {Jawad}}\ and\ \bibinfo {author} {\bibfnamefont {A.}~\bibnamefont {Iqbal}},\
  }\href {\doibase 10.1142/S0218271816500747} {\bibfield  {journal} {\bibinfo
  {journal} {Int. J. Mod. Phys. D}\ }\textbf {\bibinfo {volume} {25}},\
  \bibinfo {pages} {1650074} (\bibinfo {year} {2016})},\ \Eprint
  {http://arxiv.org/abs/1610.09961} {arXiv:1610.09961 [gr-qc]} \BibitemShut
  {NoStop}%
\bibitem [{\citenamefont {Szydlowski}\ and\ \citenamefont
  {Krawiec}(2020)}]{Szydlowski:2020ilx}%
  \BibitemOpen
  \bibfield  {author} {\bibinfo {author} {\bibfnamefont {M.}~\bibnamefont
  {Szydlowski}}\ and\ \bibinfo {author} {\bibfnamefont {A.}~\bibnamefont
  {Krawiec}},\ }\href@noop {} {\  (\bibinfo {year} {2020})},\ \Eprint
  {http://arxiv.org/abs/2006.14900} {arXiv:2006.14900 [gr-qc]} \BibitemShut
  {NoStop}%
\bibitem [{\citenamefont {Amendola}\ \emph {et~al.}(2003)\citenamefont
  {Amendola}, \citenamefont {Finelli}, \citenamefont {Burigana},\ and\
  \citenamefont {Carturan}}]{Amendola_2003}%
  \BibitemOpen
  \bibfield  {author} {\bibinfo {author} {\bibfnamefont {L.}~\bibnamefont
  {Amendola}}, \bibinfo {author} {\bibfnamefont {F.}~\bibnamefont {Finelli}},
  \bibinfo {author} {\bibfnamefont {C.}~\bibnamefont {Burigana}}, \ and\
  \bibinfo {author} {\bibfnamefont {D.}~\bibnamefont {Carturan}},\ }\href
  {\doibase 10.1088/1475-7516/2003/07/005} {\bibfield  {journal} {\bibinfo
  {journal} {Journal of Cosmology and Astroparticle Physics}\ }\textbf
  {\bibinfo {volume} {2003}},\ \bibinfo {pages} {005} (\bibinfo {year}
  {2003})}\BibitemShut {NoStop}%
\bibitem [{\citenamefont {Magana}\ \emph {et~al.}(2018)\citenamefont {Magana},
  \citenamefont {Amante}, \citenamefont {Garcia-Aspeitia},\ and\ \citenamefont
  {Motta}}]{Magana:2017nfs}%
  \BibitemOpen
  \bibfield  {author} {\bibinfo {author} {\bibfnamefont {J.}~\bibnamefont
  {Magana}}, \bibinfo {author} {\bibfnamefont {M.~H.}\ \bibnamefont {Amante}},
  \bibinfo {author} {\bibfnamefont {M.~A.}\ \bibnamefont {Garcia-Aspeitia}}, \
  and\ \bibinfo {author} {\bibfnamefont {V.}~\bibnamefont {Motta}},\ }\href
  {\doibase 10.1093/mnras/sty260} {\bibfield  {journal} {\bibinfo  {journal}
  {Mon. Not. Roy. Astron. Soc.}\ }\textbf {\bibinfo {volume} {476}},\ \bibinfo
  {pages} {1036} (\bibinfo {year} {2018})},\ \Eprint
  {http://arxiv.org/abs/1706.09848} {arXiv:1706.09848 [astro-ph.CO]}
  \BibitemShut {NoStop}%
\bibitem [{\citenamefont {Nunes}\ \emph {et~al.}(2020)\citenamefont {Nunes},
  \citenamefont {Yadav}, \citenamefont {Jesus},\ and\ \citenamefont
  {Bernui}}]{nunes:2020}%
  \BibitemOpen
  \bibfield  {author} {\bibinfo {author} {\bibfnamefont {R.~C.}\ \bibnamefont
  {Nunes}}, \bibinfo {author} {\bibfnamefont {S.~K.}\ \bibnamefont {Yadav}},
  \bibinfo {author} {\bibfnamefont {J.~F.}\ \bibnamefont {Jesus}}, \ and\
  \bibinfo {author} {\bibfnamefont {A.}~\bibnamefont {Bernui}},\ }\href@noop {}
  {\  (\bibinfo {year} {2020})},\ \Eprint {http://arxiv.org/abs/2002.09293}
  {arXiv:2002.09293 [astro-ph.CO]} \BibitemShut {NoStop}%
\bibitem [{\citenamefont {Amante}\ \emph {et~al.}(2019)\citenamefont {Amante},
  \citenamefont {Maga\~na}, \citenamefont {Motta}, \citenamefont
  {Garc\'\i{}a-Aspeitia},\ and\ \citenamefont {Verdugo}}]{Amante:2019xao}%
  \BibitemOpen
  \bibfield  {author} {\bibinfo {author} {\bibfnamefont {M.~H.}\ \bibnamefont
  {Amante}}, \bibinfo {author} {\bibfnamefont {J.}~\bibnamefont {Maga\~na}},
  \bibinfo {author} {\bibfnamefont {V.}~\bibnamefont {Motta}}, \bibinfo
  {author} {\bibfnamefont {M.~A.}\ \bibnamefont {Garc\'\i{}a-Aspeitia}}, \ and\
  \bibinfo {author} {\bibfnamefont {T.}~\bibnamefont {Verdugo}},\ }\href@noop
  {} {\  (\bibinfo {year} {2019})},\ \Eprint {http://arxiv.org/abs/1906.04107}
  {arXiv:1906.04107 [astro-ph.CO]} \BibitemShut {NoStop}%
\bibitem [{\citenamefont {{Ch{\'a}vez}}\ \emph {et~al.}(2012)\citenamefont
  {{Ch{\'a}vez}}, \citenamefont {{Terlevich}}, \citenamefont {{Terlevich}},
  \citenamefont {{Plionis}}, \citenamefont {{Bresolin}}, \citenamefont
  {{Basilakos}},\ and\ \citenamefont {{Melnick}}}]{Chavez2012}%
  \BibitemOpen
  \bibfield  {author} {\bibinfo {author} {\bibfnamefont {R.}~\bibnamefont
  {{Ch{\'a}vez}}}, \bibinfo {author} {\bibfnamefont {E.}~\bibnamefont
  {{Terlevich}}}, \bibinfo {author} {\bibfnamefont {R.}~\bibnamefont
  {{Terlevich}}}, \bibinfo {author} {\bibfnamefont {M.}~\bibnamefont
  {{Plionis}}}, \bibinfo {author} {\bibfnamefont {F.}~\bibnamefont
  {{Bresolin}}}, \bibinfo {author} {\bibfnamefont {S.}~\bibnamefont
  {{Basilakos}}}, \ and\ \bibinfo {author} {\bibfnamefont {J.}~\bibnamefont
  {{Melnick}}},\ }\href {\doibase 10.1111/j.1745-3933.2012.01299.x} {\bibfield
  {journal} {\bibinfo  {journal} {MNRAS}\ }\textbf {\bibinfo {volume} {425}},\
  \bibinfo {pages} {L56} (\bibinfo {year} {2012})},\ \Eprint
  {http://arxiv.org/abs/1203.6222} {arXiv:1203.6222 [astro-ph.CO]} \BibitemShut
  {NoStop}%
\bibitem [{\citenamefont {{Ch{\'a}vez}}\ \emph {et~al.}(2014)\citenamefont
  {{Ch{\'a}vez}}, \citenamefont {{Terlevich}}, \citenamefont {{Terlevich}},
  \citenamefont {{Bresolin}}, \citenamefont {{Melnick}}, \citenamefont
  {{Plionis}},\ and\ \citenamefont {{Basilakos}}}]{Chavez2014}%
  \BibitemOpen
  \bibfield  {author} {\bibinfo {author} {\bibfnamefont {R.}~\bibnamefont
  {{Ch{\'a}vez}}}, \bibinfo {author} {\bibfnamefont {R.}~\bibnamefont
  {{Terlevich}}}, \bibinfo {author} {\bibfnamefont {E.}~\bibnamefont
  {{Terlevich}}}, \bibinfo {author} {\bibfnamefont {F.}~\bibnamefont
  {{Bresolin}}}, \bibinfo {author} {\bibfnamefont {J.}~\bibnamefont
  {{Melnick}}}, \bibinfo {author} {\bibfnamefont {M.}~\bibnamefont
  {{Plionis}}}, \ and\ \bibinfo {author} {\bibfnamefont {S.}~\bibnamefont
  {{Basilakos}}},\ }\href {\doibase 10.1093/mnras/stu987} {\bibfield  {journal}
  {\bibinfo  {journal} {MNRAS}\ }\textbf {\bibinfo {volume} {442}},\ \bibinfo
  {pages} {3565} (\bibinfo {year} {2014})},\ \Eprint
  {http://arxiv.org/abs/1405.4010} {arXiv:1405.4010 [astro-ph.GA]} \BibitemShut
  {NoStop}%
\bibitem [{\citenamefont {{Terlevich}}\ \emph {et~al.}(2015)\citenamefont
  {{Terlevich}}, \citenamefont {{Terlevich}}, \citenamefont {{Melnick}},
  \citenamefont {{Ch{\'a}vez}}, \citenamefont {{Plionis}}, \citenamefont
  {{Bresolin}},\ and\ \citenamefont {{Basilakos}}}]{Terlevich2015}%
  \BibitemOpen
  \bibfield  {author} {\bibinfo {author} {\bibfnamefont {R.}~\bibnamefont
  {{Terlevich}}}, \bibinfo {author} {\bibfnamefont {E.}~\bibnamefont
  {{Terlevich}}}, \bibinfo {author} {\bibfnamefont {J.}~\bibnamefont
  {{Melnick}}}, \bibinfo {author} {\bibfnamefont {R.}~\bibnamefont
  {{Ch{\'a}vez}}}, \bibinfo {author} {\bibfnamefont {M.}~\bibnamefont
  {{Plionis}}}, \bibinfo {author} {\bibfnamefont {F.}~\bibnamefont
  {{Bresolin}}}, \ and\ \bibinfo {author} {\bibfnamefont {S.}~\bibnamefont
  {{Basilakos}}},\ }\href {\doibase 10.1093/mnras/stv1128} {\bibfield
  {journal} {\bibinfo  {journal} {MNRAS}\ }\textbf {\bibinfo {volume} {451}},\
  \bibinfo {pages} {3001} (\bibinfo {year} {2015})},\ \Eprint
  {http://arxiv.org/abs/1505.04376} {arXiv:1505.04376 [astro-ph.CO]}
  \BibitemShut {NoStop}%
\bibitem [{\citenamefont {{Ch{\'a}vez}}\ \emph {et~al.}(2016)\citenamefont
  {{Ch{\'a}vez}}, \citenamefont {{Plionis}}, \citenamefont {{Basilakos}},
  \citenamefont {{Terlevich}}, \citenamefont {{Terlevich}}, \citenamefont
  {{Melnick}}, \citenamefont {{Bresolin}},\ and\ \citenamefont
  {{Gonz{\'a}lez-Mor{\'a}n}}}]{Chavez2016}%
  \BibitemOpen
  \bibfield  {author} {\bibinfo {author} {\bibfnamefont {R.}~\bibnamefont
  {{Ch{\'a}vez}}}, \bibinfo {author} {\bibfnamefont {M.}~\bibnamefont
  {{Plionis}}}, \bibinfo {author} {\bibfnamefont {S.}~\bibnamefont
  {{Basilakos}}}, \bibinfo {author} {\bibfnamefont {R.}~\bibnamefont
  {{Terlevich}}}, \bibinfo {author} {\bibfnamefont {E.}~\bibnamefont
  {{Terlevich}}}, \bibinfo {author} {\bibfnamefont {J.}~\bibnamefont
  {{Melnick}}}, \bibinfo {author} {\bibfnamefont {F.}~\bibnamefont
  {{Bresolin}}}, \ and\ \bibinfo {author} {\bibfnamefont {A.~L.}\ \bibnamefont
  {{Gonz{\'a}lez-Mor{\'a}n}}},\ }\href {\doibase 10.1093/mnras/stw1813}
  {\bibfield  {journal} {\bibinfo  {journal} {MNRAS}\ }\textbf {\bibinfo
  {volume} {462}},\ \bibinfo {pages} {2431} (\bibinfo {year} {2016})},\ \Eprint
  {http://arxiv.org/abs/1607.06458} {arXiv:1607.06458 [astro-ph.CO]}
  \BibitemShut {NoStop}%
\bibitem [{\citenamefont {{Gonz{\'a}lez-Mor{\'a}n}}\ \emph
  {et~al.}(2019)\citenamefont {{Gonz{\'a}lez-Mor{\'a}n}}, \citenamefont
  {{Ch{\'a}vez}}, \citenamefont {{Terlevich}}, \citenamefont {{Terlevich}},
  \citenamefont {{Bresolin}}, \citenamefont {{Fern{\'a}ndez-Arenas}},
  \citenamefont {{Plionis}}, \citenamefont {{Basilakos}}, \citenamefont
  {{Melnick}},\ and\ \citenamefont {{Telles}}}]{GonzalezMoran2019}%
  \BibitemOpen
  \bibfield  {author} {\bibinfo {author} {\bibfnamefont {A.~L.}\ \bibnamefont
  {{Gonz{\'a}lez-Mor{\'a}n}}}, \bibinfo {author} {\bibfnamefont
  {R.}~\bibnamefont {{Ch{\'a}vez}}}, \bibinfo {author} {\bibfnamefont
  {R.}~\bibnamefont {{Terlevich}}}, \bibinfo {author} {\bibfnamefont
  {E.}~\bibnamefont {{Terlevich}}}, \bibinfo {author} {\bibfnamefont
  {F.}~\bibnamefont {{Bresolin}}}, \bibinfo {author} {\bibfnamefont
  {D.}~\bibnamefont {{Fern{\'a}ndez-Arenas}}}, \bibinfo {author} {\bibfnamefont
  {M.}~\bibnamefont {{Plionis}}}, \bibinfo {author} {\bibfnamefont
  {S.}~\bibnamefont {{Basilakos}}}, \bibinfo {author} {\bibfnamefont
  {J.}~\bibnamefont {{Melnick}}}, \ and\ \bibinfo {author} {\bibfnamefont
  {E.}~\bibnamefont {{Telles}}},\ }\href {\doibase 10.1093/mnras/stz1577}
  {\bibfield  {journal} {\bibinfo  {journal} {MNRAS}\ }\textbf {\bibinfo
  {volume} {487}},\ \bibinfo {pages} {4669} (\bibinfo {year} {2019})},\ \Eprint
  {http://arxiv.org/abs/1906.02195} {arXiv:1906.02195 [astro-ph.GA]}
  \BibitemShut {NoStop}%
\bibitem [{\citenamefont {Sahni}\ \emph {et~al.}(2008)\citenamefont {Sahni},
  \citenamefont {Shafieloo},\ and\ \citenamefont {Starobinsky}}]{Sanhi:2008}%
  \BibitemOpen
  \bibfield  {author} {\bibinfo {author} {\bibfnamefont {V.}~\bibnamefont
  {Sahni}}, \bibinfo {author} {\bibfnamefont {A.}~\bibnamefont {Shafieloo}}, \
  and\ \bibinfo {author} {\bibfnamefont {A.~A.}\ \bibnamefont {Starobinsky}},\
  }\href {\doibase 10.1103/PhysRevD.78.103502} {\bibfield  {journal} {\bibinfo
  {journal} {Phys. Rev. D}\ }\textbf {\bibinfo {volume} {78}},\ \bibinfo
  {pages} {103502} (\bibinfo {year} {2008})}\BibitemShut {NoStop}%
\bibitem [{\citenamefont {Foreman-Mackey}\ \emph {et~al.}(2013)\citenamefont
  {Foreman-Mackey}, \citenamefont {Hogg}, \citenamefont {Lang},\ and\
  \citenamefont {Goodman}}]{Foreman:2013}%
  \BibitemOpen
  \bibfield  {author} {\bibinfo {author} {\bibfnamefont {D.}~\bibnamefont
  {Foreman-Mackey}}, \bibinfo {author} {\bibfnamefont {D.~W.}\ \bibnamefont
  {Hogg}}, \bibinfo {author} {\bibfnamefont {D.}~\bibnamefont {Lang}}, \ and\
  \bibinfo {author} {\bibfnamefont {J.}~\bibnamefont {Goodman}},\ }\href
  {\doibase 10.1086/670067} {\bibfield  {journal} {\bibinfo  {journal}
  {Publications of the Astronomical Society of the Pacific}\ }\textbf {\bibinfo
  {volume} {125}},\ \bibinfo {pages} {306} (\bibinfo {year}
  {2013})}\BibitemShut {NoStop}%
\bibitem [{\citenamefont {Gelman}\ and\ \citenamefont
  {Rubin}(1992)}]{Gelman:1992}%
  \BibitemOpen
  \bibfield  {author} {\bibinfo {author} {\bibfnamefont {A.}~\bibnamefont
  {Gelman}}\ and\ \bibinfo {author} {\bibfnamefont {D.}~\bibnamefont {Rubin}},\
  }\href {\doibase 10.1103/PhysRevD.67.101301} {\bibfield  {journal} {\bibinfo
  {journal} {Statistical Science}\ }\textbf {\bibinfo {volume} {67}},\ \bibinfo
  {pages} {457} (\bibinfo {year} {1992})}\BibitemShut {NoStop}%
\bibitem [{\citenamefont {Cao}\ \emph {et~al.}(2020)\citenamefont {Cao},
  \citenamefont {Ryan},\ and\ \citenamefont {Ratra}}]{Cao:2020jgu}%
  \BibitemOpen
  \bibfield  {author} {\bibinfo {author} {\bibfnamefont {S.}~\bibnamefont
  {Cao}}, \bibinfo {author} {\bibfnamefont {J.}~\bibnamefont {Ryan}}, \ and\
  \bibinfo {author} {\bibfnamefont {B.}~\bibnamefont {Ratra}},\ }\href
  {\doibase 10.1093/mnras/staa2190} {\bibfield  {journal} {\bibinfo  {journal}
  {Mon. Not. Roy. Astron. Soc.}\ }\textbf {\bibinfo {volume} {497}},\ \bibinfo
  {pages} {3191} (\bibinfo {year} {2020})},\ \Eprint
  {http://arxiv.org/abs/2005.12617} {arXiv:2005.12617 [astro-ph.CO]}
  \BibitemShut {NoStop}%
\bibitem [{\citenamefont {Akaike}(1974)}]{AIC:1974}%
  \BibitemOpen
  \bibfield  {author} {\bibinfo {author} {\bibfnamefont {H.}~\bibnamefont
  {Akaike}},\ }\href {\doibase 10.1109/TAC.1974.1100705} {\bibfield  {journal}
  {\bibinfo  {journal} {IEEE Transactions on Automatic Control}\ }\textbf
  {\bibinfo {volume} {19}},\ \bibinfo {pages} {716} (\bibinfo {year}
  {1974})}\BibitemShut {NoStop}%
\bibitem [{\citenamefont {Sugiura}(1978)}]{Sugiura:1978}%
  \BibitemOpen
  \bibfield  {author} {\bibinfo {author} {\bibfnamefont {N.}~\bibnamefont
  {Sugiura}},\ }\href {\doibase 10.1080/03610927808827599} {\bibfield
  {journal} {\bibinfo  {journal} {Communications in Statistics - Theory and
  Methods}\ }\textbf {\bibinfo {volume} {7}},\ \bibinfo {pages} {13} (\bibinfo
  {year} {1978})}\BibitemShut {NoStop}%
\bibitem [{\citenamefont {Hurvich}\ and\ \citenamefont
  {Tsai}(1989)}]{AICc:1989}%
  \BibitemOpen
  \bibfield  {author} {\bibinfo {author} {\bibfnamefont {C.~M.}\ \bibnamefont
  {Hurvich}}\ and\ \bibinfo {author} {\bibfnamefont {C.~L.}\ \bibnamefont
  {Tsai}},\ }\href@noop {} {\bibfield  {journal} {\bibinfo  {journal}
  {Biometrika}\ }\textbf {\bibinfo {volume} {76}},\ \bibinfo {pages} {297}
  (\bibinfo {year} {1989})}\BibitemShut {NoStop}%
\bibitem [{\citenamefont {Schwarz}(1978)}]{schwarz1978}%
  \BibitemOpen
  \bibfield  {author} {\bibinfo {author} {\bibfnamefont {G.}~\bibnamefont
  {Schwarz}},\ }\href {\doibase 10.1214/aos/1176344136} {\bibfield  {journal}
  {\bibinfo  {journal} {Ann. Statist.}\ }\textbf {\bibinfo {volume} {6}},\
  \bibinfo {pages} {461} (\bibinfo {year} {1978})}\BibitemShut {NoStop}%
\bibitem [{\citenamefont {Herrera-Zamorano}\ \emph {et~al.}(2020)\citenamefont
  {Herrera-Zamorano}, \citenamefont {Hern\'andez-Almada},\ and\ \citenamefont
  {Garc\'ia-Aspeitia}}]{Herrera:2020}%
  \BibitemOpen
  \bibfield  {author} {\bibinfo {author} {\bibfnamefont {L.}~\bibnamefont
  {Herrera-Zamorano}}, \bibinfo {author} {\bibfnamefont {A.}~\bibnamefont
  {Hern\'andez-Almada}}, \ and\ \bibinfo {author} {\bibfnamefont
  {M.}~\bibnamefont {Garc\'ia-Aspeitia}},\ }\href {\doibase
  10.1140/epjc/s10052-020-8225-y} {\bibfield  {journal} {\bibinfo  {journal}
  {Eur. Phys. J. C}\ }\textbf {\bibinfo {volume} {80}},\ \bibinfo {pages} {637}
  (\bibinfo {year} {2020})},\ \Eprint {http://arxiv.org/abs/2007.04507}
  {arXiv:2007.04507} \BibitemShut {NoStop}%
\bibitem [{\citenamefont {Normann}\ and\ \citenamefont
  {Brevik}(2016)}]{Normann:2016}%
  \BibitemOpen
  \bibfield  {author} {\bibinfo {author} {\bibfnamefont {B.~D.}\ \bibnamefont
  {Normann}}\ and\ \bibinfo {author} {\bibfnamefont {I.}~\bibnamefont
  {Brevik}},\ }\href {\doibase 10.3390/e18060215} {\bibfield  {journal}
  {\bibinfo  {journal} {Entropy}\ }\textbf {\bibinfo {volume} {18}} (\bibinfo
  {year} {2016}),\ 10.3390/e18060215}\BibitemShut {NoStop}%
\bibitem [{\citenamefont {{Riess}}\ \emph {et~al.}(2019)\citenamefont
  {{Riess}}, \citenamefont {{Casertano}}, \citenamefont {{Yuan}}, \citenamefont
  {{Macri}},\ and\ \citenamefont {{Scolnic}}}]{Riess:2019}%
  \BibitemOpen
  \bibfield  {author} {\bibinfo {author} {\bibfnamefont {A.~G.}\ \bibnamefont
  {{Riess}}}, \bibinfo {author} {\bibfnamefont {S.}~\bibnamefont
  {{Casertano}}}, \bibinfo {author} {\bibfnamefont {W.}~\bibnamefont {{Yuan}}},
  \bibinfo {author} {\bibfnamefont {L.~M.}\ \bibnamefont {{Macri}}}, \ and\
  \bibinfo {author} {\bibfnamefont {D.}~\bibnamefont {{Scolnic}}},\ }\href@noop
  {} {\bibfield  {journal} {\bibinfo  {journal} {arXiv e-prints}\ } (\bibinfo
  {year} {2019})},\ \Eprint {http://arxiv.org/abs/1903.07603}
  {arXiv:1903.07603} \BibitemShut {NoStop}%
\end{thebibliography}%

\end{document}